\documentclass[useAMS,usenatbib,usegraphicx,a4]{mn2e}
\usepackage{times}
\usepackage{url}
\usepackage{epsfig}
\usepackage[flushleft]{threeparttable}
\usepackage{amssymb}
\usepackage{amsmath}
\usepackage{pdfpages}
\usepackage{array}
\usepackage{comment}

\newcolumntype{H}{>{\setbox0=\hbox\bgroup}c<{\egroup}@{}}

\def\apj{ApJ}
\def\apjs{ApJS}
\def\mnras{MNRAS}

\def\aj{AJ}

\title[SkyMapper colours of Seyfert galaxies]{SkyMapper colours of Seyfert galaxies and Changing-Look AGN}

\author[Wolf et al.]{Christian Wolf$^{1,2}$, Jacob Golding$^{1}$, Wei Jeat Hon$^3$, and Christopher A. Onken$^{1,2}$ \\
$^1$Research School of Astronomy and Astrophysics, Australian National University, Canberra ACT 2611, Australia, E-mail: christian.wolf@anu.edu.au\\
$^2$Centre for Gravitational Astrophysics, Australian National University, Canberra ACT 2611, Australia\\
$^3$School of Physics, University of Melbourne, Parkville, Victoria 3010, Australia \\
}
\newcommand{\refbf}{} 

\begin{document}

\date{Accepted 2020 September 9. Received 2020 September 8; in original form 2020 July 6.}
\maketitle

\begin{abstract}
We study the utility of broad-band colours in the SkyMapper Southern Survey for selecting Seyfert galaxies at low luminosity. We find that the $u-v$ index, built from the ultraviolet $u$ and violet $v$ filters, separates normal galaxies, starburst galaxies and type-1 AGN. This $u-v$ index is not sensitive to age or metallicity in a stellar population but is instead a quenching-and-bursting indicator in galaxies and detects power-law continua in type-1 AGN. Using over 25,000 galaxies at $z<0.1$ from 6dFGS, we find a selection cut based on $u-v$ and central $u$ band brightness that identifies type-1 AGN. By eyeballing 6dFGS spectra we classify new Seyfert galaxies of type 1 to 1.8. Our sample includes eight known Changing-Look AGN, two of which show such strong variability that they move across the selection cut during the five years of SkyMapper observations in DR3, along mixing sequences of nuclear and host galaxy light. {\refbf We identify 46 Changing-Look AGN candidates in our sample, one of which has been reported as a type-IIn supernova. We show that this transient persists for at least five years and marks a flare in a Seyfert-1 period of a new Changing-Look AGN}. 
\end{abstract}

\section{Introduction}\label{sec:intro}

Broad-band spectral energy distributions (SEDs), a.k.a. colours, of galaxies have often been used to classify galaxies or characterise their stellar populations \citep[see e.g. review by][]{Walcher11}. Properties that are evident in SEDs include star formation rate and dust content \citep[e.g.][]{daCunha08} as well as non-stellar radiation from Active Galactic Nuclei \cite[AGN; see reviews by][]{Weedman77,Osterbrock84,VCV00}.  

Colours also proved to be a powerful tool for finding AGN, which has been exploited for decades now \citep[see reviews by][]{HewettFoltz94, Osmer04}. UV-optical colours reveal broad-line AGN, whose featureless continua alter the UV-optical SEDs of Seyfert galaxies with moderate luminosity and totally dominate the SEDs of powerful quasi-stellar objects (QSOs). Some methods achieve higher completeness than others, e.g. by using objective-prism spectroscopy \citep{Wisotzki00} or medium-band filters \citep{Wolf03} to detect the broad emission lines in better-resolved SEDs. 
Other wavelengths helped finding AGN with less obvious continua: moderately dust-reddenend AGN, e.g., were better found when \citet{WHF00} included near-IR photometry in their 'KX method', which compares the power-law nature of AGN SEDs against the convex spectra of stars. Further into the mid-IR, this idea helped the discovery of high-luminosity QSOs \citep[e.g.][]{Lacy04,Stern05,Richards06,Wu12}. However, using mid-IR data 
\citet{Asmus20} identify new low-luminosity AGNs even within a 100~Mpc neighbourhood of our Milky Way, where new identifications may not have been expected. Finally, X-rays revealed optically obscured systems and characterised their obscuration in more detail \citep[see review by][]{BrandtHasinger05}. 

This paper focuses on Seyfert galaxies, which are AGN with lower luminosity than the powerful QSOs that are easily observed all the way to the edge of the Universe \citep[e.g.][]{Fan01}. Irrespective of observing wavelength, the contrast between nucleus and host galaxy is an issue for active nuclei of lower luminosity \citep[see][and follow-up papers]{Ho95} and depends on spatial resolution --  both, finding AGN using SEDs or classifying their spectra is challenging, when the change {\refbf in spectrum or SED caused} by the nucleus is hardly above the noise. 

Monitoring the sky for variable objects provided a separate approach to identify AGN, at least those whose usually variable and UV-optically luminous accretion disk was not obscured by dust \citep[e.g.][]{Hawkins93,Graham17}. Variability was recognised as ubiquitous long ago, but it was long assumed that it could not be strong enough to change profoundly the spectral type of an AGN \citep[and references therein]{Lawrence18}. When \citet{TO76} found in the Seyfert-1 galaxy NGC~7603 that the broad H$\beta$ emission line disappeared and reappeared within a couple of years, it seemed to be an outlier phenomenon, perhaps best explained by dust extinction varying with time. As more cases were observed \citep[e.g.][]{Alloin86,Cohen86,Aretxaga99} they were labelled Changing-Look AGN \citep[or CLAGN,][]{Denney14,Shappee14} to describe major changes in their spectral type \citep[as defined e.g. in][]{Osterbrock81}. 

It is now becoming evident that drastic changes in the appearance of AGN emission lines are remarkably common. This implies that the SED of an AGN is a temporary and variable characteristic, and surveys conducted at different times might find different sets of objects. Thus, time-domain studies in large data sets open up an unexpected dimension in AGN research. Whether all AGN are potential CLAGN just observed in one phase or another, or there is a category of either AGN or host galaxies, whose physical conditions make them prone to major change, is not clear at this stage.

Several groups discovered multiple CLAGN by comparing spectra of AGN with repeat observations at different epochs \citep[e.g.][]{MacLeod16,Yang18,Hon20}. \citet{Wolf18} found three CLAGN by searching for AGN in the Hamburg-ESO QSO Survey \citep{Wisotzki00}, whose brightness had declined by the time they were observed in the SkyMapper Southern Survey. CLAGN rates can be estimated from follow-up of complete AGN samples: e.g. more complete follow-up will investigate the rate of turn-off changes from the type-1 sample of the Hamburg-ESO Survey, which is regarded to be exceptionally complete. Conversely, the rate of turn-on events can be studied by monitoring a complete sample of type-2 AGN for brightening and the appearance of type-1 SEDs.

In this paper, we return to identifying type-1 AGN from UV-optical colours. We use large catalogues of low-redshift galaxies in conjunction with broad-band SEDs observed by the SkyMapper Southern Survey \citep{Wolf18,Onken19} to identify which Southern galaxies appear to be Seyfert-1 types now that they were imaged by SkyMapper. We can then inspect older spectra of these candidate Sy-1 objects, and when those are more consistent with a type-2 AGN, the galaxy {is a turn-on} CLAGN candidate. {\refbf Conversely, we can also identify turn-off CL-AGN candidates as objects that combine earlier Seyfert-1 spectra with SkyMapper colours that are untypical for Seyfert-1 galaxies.} We will test this approach with existing findings on known CLAGN in the Southern hemisphere.

The aim of distinguishing Seyfert-1 galaxies from non-AGN requires us to investigate more widely the colours of low-redshift galaxies in SkyMapper, and especially the colour space occupied by starburst galaxies. There is general interest in this exploration as there are many galaxies in the Southern sky, which have not been studied in as much detail as their Northern cousins that were imaged a long time ago by the Sloan Digital Sky Survey \citep[SDSS;][]{York00}. The Southern hemisphere is now not only being covered with SkyMapper as an optical multi-band survey, but will soon be covered by the HI survey WALLABY \citep{Koribalski20} that is created with the Australian SKA Pathfinder \citep[ASKAP;][]{Jo07} facility. Combining optical and HI data is of obvious interest for studying the cycle of gas and star formation in galaxies in spatially resolved form, as the hemisphere contains over 10,000 galaxies that are larger than one arcminute in diameter. 

In this paper, we explore the colours of low-redshift galaxies in the SkyMapper Southern Survey Data Release 3 (DR3). We focus on the value of its unique filters $u$ and $v$, which differ from the SDSS filter set used by most modern large-area surveys; this part is currently unexplored in the literature as are potential uses of the $uv$ colour for classifying and characterising galaxies. We look in particular at differences between starburst galaxies and AGN as well as AGN of different type, with an outlook to finding CLAGN.

\begin{figure}
\begin{center}
\includegraphics[angle=270,width=\columnwidth]{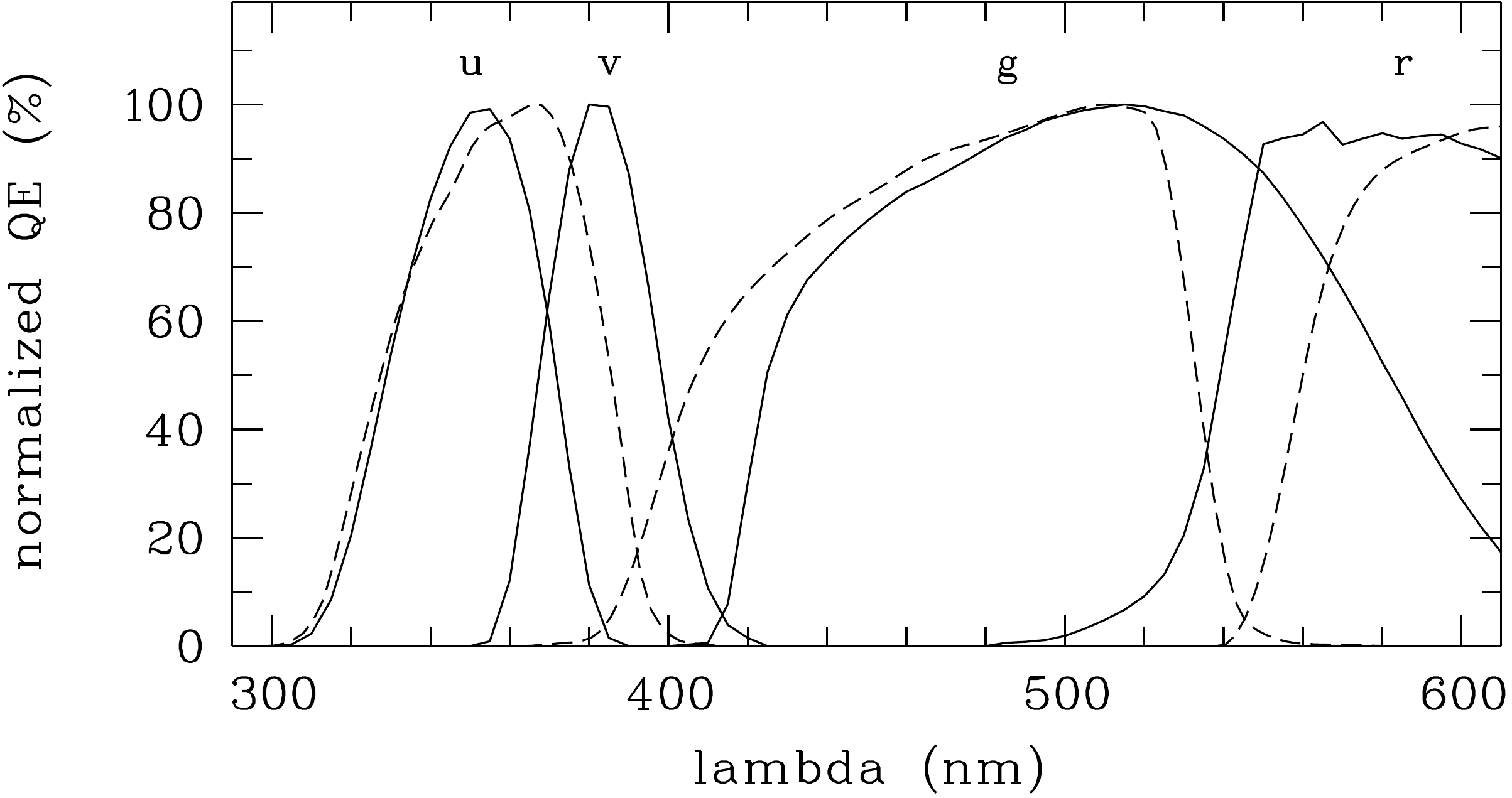}
\caption{\refbf Comparison of normalised efficiency curves for SDSS $ug$ bands (dashed lines) and SkyMapper $uvg$ bands (solid lines, $v$ = violet).}\label{uvg}
\end{center}
\end{figure}

In Sect.~2 we consider the SkyMapper filter set and how it is different from SDSS filters and then model galaxy and AGN colours through population synthesis and AGN templates. Sect.~3 presents the colours of low-redshift galaxies as observed in DR3. In Sect.~4 we study the separation of active and inactive galaxies and sub-types of AGN as well as the mixing sequence of colours from pure stellar populations to pure AGN. In Sect.~5 we discuss the known CLAGN in the SkyMapper sample and in Sect.~6 we conclude. Throughout the paper we use AB magnitudes and a Hubble-Lema\^itre constant of $H_0=70$~km~sec$^{-1}$~Mpc$^{-1}$.

\section{Expectations for S{\sevensize\bf ky}M{\sevensize\bf apper} colours}

In this section, we describe the differences between the SkyMapper filter set and the more common filter sets derived from the SDSS $ugriz$ system. Of particular interest in SkyMapper are the bespoke ultraviolet $u$ and violet $v$ bands. The filter curves were published by \citet{Bessell11}. We use synthetic photometry to model unobscured Seyfert nuclei from SED grids based on standard QSO templates and galaxy SEDs from stellar population synthesis.

\subsection{The SkyMapper filter set: from stars to galaxies}

When SkyMapper was planned, the Sloan Digital Sky Survey had already revolutionised galaxy surveys by observing much of the Northern sky. As SkyMapper filters were designed, their scientific niche was seen to be improved studies of Milky Way stars. 

The biggest difference between SkyMapper and SDSS filters is on the blue side of 400~nm (see Table~\ref{filtab} {\refbf and Fig.~\ref{uvg}}): where SDSS has a $u$ band with $(\lambda_{\rm cen}/{\rm FWHM})=(358{\rm nm}/55{\rm nm})$ SkyMapper has a violet $v$ band (384/28) and a more ultraviolet $u$ band (349/42). This choice provides photometric sensitivity to the surface gravity of stars as the strength of the Hydrogen Balmer break affects the $u-v$ colour, as well as to metallicity as metal lines affect the $v-g$ colour. By exploiting this information, SkyMapper has facilitated the discovery of a large sample of extremely metal-poor stars \citep{DaCosta19} including the most chemically pristine star currently known \citep{Keller14}. In this paper we demonstrate the use of this filter pair in near-field extragalactic work.

Filters that are useful for characterising stars should also help with characterising stellar populations in galaxies at redshifts close to zero. SEDs of more distant galaxies are of course redshifted out of sync with the passbands. We first note that the SkyMapper colour index $u-v$ takes on its reddest value for main-sequence A-type stars. We then expect the $u-v$ colour of a stellar population to depend on the fraction of A-type stars in the mix. After a quenching event, the young O- and B-type stars die away and leave the blue side of the galaxy SED dominated by the bright $(u-v)$-red A-stars. Conversely, in the event of a starburst a rapid increase in the fraction of O- and B-stars above typical levels makes the $u-v$ index bluer than usual. A continuously star-forming and steadily evolving population, however, has a constant O-to-A star ratio and will thus show a constant $u-v$ index, irrespective of the detailed history of an older underlying population. 

In the following, we explore in detail how the $u-v$ colour responds to temporal changes in the star-formation rate of a galaxy and to AGN contributions. Colour indices made from $griz$ bands are expected to behave similar to the well-studied SDSS colours. 

\begin{table}
\caption{SkyMapper filters and DR3 images: FWHM is for the median PSF. 10$\sigma$ limit is the mean 5$\arcsec$-aperture magnitude of 10$\sigma$ sources in single Main Survey exposures. $R$ is the extinction coefficient from \citet{Wolf18}, to be used with $A_{\rm band}=R_{\rm band}\times E(B-V)_{\rm SFD}$.}
\label{filtab}      
\centering          
\begin{tabular}{llccccc}
\hline       
Filter & $\lambda_{\rm cen}$/$\Delta \lambda$ & FWHM & $10\sigma$ limit & $R$ \\ 
	   & (nm) & (arcsec) & (ABmag) \\
\hline
$u$ & 349/42  & $3.1$ & 19.12 & 4.294 \\ 
$v$ & 384/28  & $2.9$ & 19.26 & 4.026 \\ 
$g$ & 510/156 & $2.6$ & 20.83 & 2.986 \\ 
$r$ & 617/156 & $2.4$ & 20.43 & 2.288 \\ 
$i$ & 779/140 & $2.3$ & 19.45 & 1.588 \\ 
$z$ & 916/84  & $2.3$ & 18.69 & 1.206 \\ 
\hline                  
\end{tabular}
\end{table}

\subsection{Stellar population synthesis}

The aim of this step is to see what part of optical colour space may be expected to be occupied by galaxies considering a broad range of evolutionary options. It is not to interpret star-formation histories in the face of degeneracies in optical photometry. We target galaxies in the local ($z\la 0.1$) Universe, which may appear young or old, but often include a substantial amount of old or intermediate-age stars in either case. For stellar populations undergoing only secular evolution we assume no particular formation redshift and consider a large range of possibilities in the star-formation history. Mature galaxies might still undergo quenching or bursting events long after most of the stars in a galaxy have formed. To capture this, we model stellar populations with exponential star-formation histories proceeding for 10~Gyr, before a quenching or burst event happens. The choice of 10~Gyr is degenerate with other parameters of the evolution and hence not critical for predicting galaxy SEDs in a small number of bands. Immediately following the event, the SEDs of the populations change rapidly so we need to calculate their SED evolution with fine time resolution. 

For population synthesis we use the {\it Stochastically Lighting Up Galaxies} (SLUG) package by \citet{SLUG}. In the long term, we intend to study the influence of stochasticity when considering small parts of well-resolved galaxies, possibly even at low surface brightness. In this first paper on galaxy colours in SkyMapper, however, we consider massive populations with little influence from stochasticity. Hence, we did not exploit SLUG's capability to follow each single star in a large population and modified our approach to speed it up: we first run SLUG for 5~Myr with a constant star formation rate to create an elementary building block of a star-formation history with a total stellar mass of $10^5{\rm M}_\odot$. We sample the SED evolution of this building block at a high time resolution of 10~Myr for a duration of 14~Gyr. In a second step, we combine the building blocks of a more complex star formation history into a full SED. The main disadvantage of our method relative to the original SLUG mode is that we do not track metallicity evolution and instead assume a solar metallicity throughout. This choice affects mostly the age-tagging of old populations given the well-known age-metallicity degeneracy in broad-band SEDs \citep{Worthey94}, but it does not shrink the space occupied by our models.

\begin{figure}
\begin{center}
\includegraphics[angle=270,width=0.9\columnwidth]{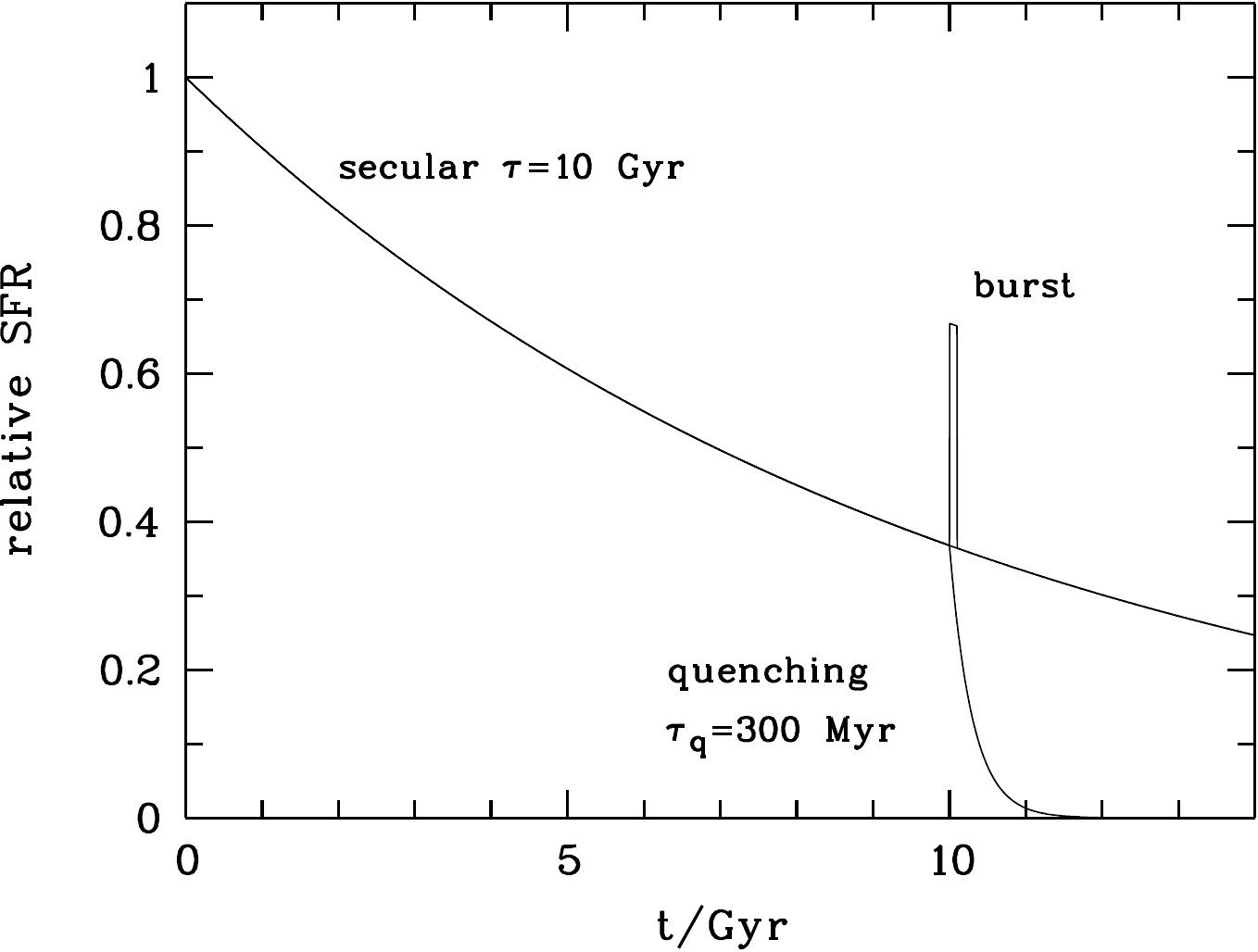}
\caption{We use three types of star formation history (SFH): (1) a secular evolution without any special event, (2) an SFH permanently quenched after 10~Gyr, and (3) an SFH with a temporary burst. Three examples are shown for an underlying secular SFH with a decline time scale of $\tau=10$~Gyr.}\label{SFH_example}
\end{center}
\end{figure}

We consider a range of SFHs with a parametrised grid, where the principal stellar population is always described as an exponential model, $SFR(t)=Ce^{-t/\tau}$, with
$\tau$ ranging from 3 to 100~Gyr. 
At $t=10$~Gyr, we place an event that is either a starburst or a quenching event. Quenching is parametrised by a second tau model with $\tau_q \in (0.03, 0.1, 0.3)$~Gyr. We also use $\tau_q=0$ for sudden complete quenching. Bursts are modelled with additional star formation that remains constant for 100~Myr and ranges from 1\% to 100\% of the instantaneous SFR at $t=0$. Figure~\ref{SFH_example} shows a sketch of this SFH parametrisation.

\begin{figure*}
\begin{center}
\includegraphics[width=\textwidth]{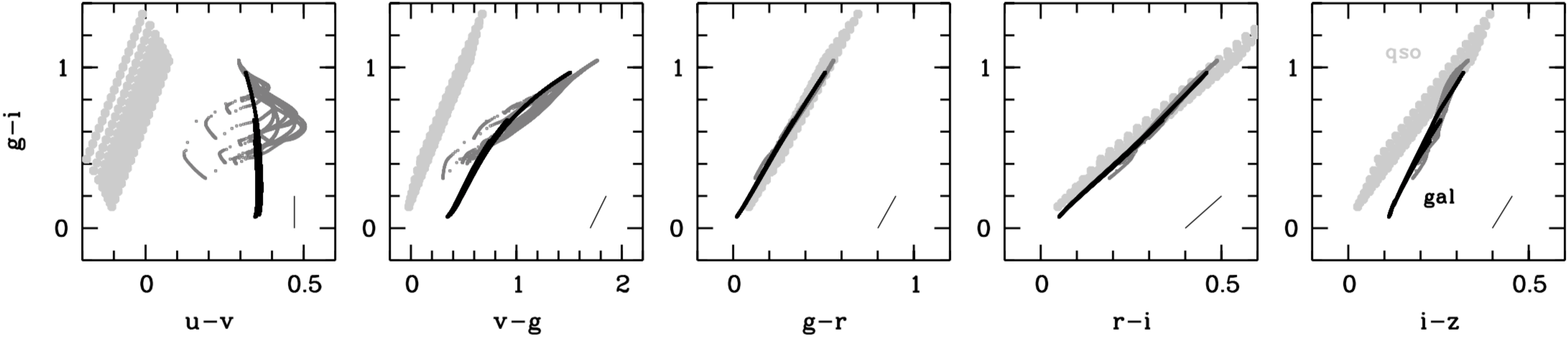}
\caption{Colour locus of QSOs (light grey) and galaxies for a variety of evolutionary scenarios including secular evolution for 14~Gyr (black), quenching and starburst events (dark grey). Galaxies form a 1-parameter family in $griz$ irrespective of their star-formation history and dust extinction (see dust vector on the bottom right in each panel). The $u$ and $v$ bands both add a dimension to the galaxy family. Outside a possible active nucleus of a galaxy, the $u-v$ index is effectively a burst-and-quench indicator, and within a nucleus it can be an AGN indicator, both provided that the galaxy redshift remains close to 0. }\label{CC}
\end{center}
\end{figure*}

\subsection{The galaxy locus in SkyMapper colours}

Figure~\ref{CC} shows the SkyMapper colour locus at redshift 0, for the grid of galaxy SEDs described above, over a time range from 1 to 14~Gyr after formation, and assuming no dust reddening.
The three panels on the right show that galaxy colours in the passbands $griz$ span effectively only a 1-parameter family of colours, where time since formation, the $\tau$ of secular evolution, and the strength of and phase within a quenching or bursting event are all degenerate. The dust vector in the figure shows that even dust reddens the SEDs along the same 1-parameter relation. In the SDSS system, a similar tight relation prevails, albeit shifted slightly due to differences between the two filter sets.

The left panels of Fig.~\ref{CC} show information added by filters on the blue side of 400~nm. The $v-g$ index straddles the 4000\AA -break and is thus close to a traditional $D_{4000}$ measurement, which is mostly driven by age of the stellar population or specific star-formation rate. It shows the effects of a second parameter in the galaxy SED family, driven by quenching and bursting events. As is well known from interpretations of colour-magnitude relations, the colour is also sensitive to metallicity in old stellar populations \citep[e.g.][]{Bower92,Poggianti97}, which is not explored further in this work. 

The left-most panel shows the $u-v$ index, which hardly depends on $\tau$, or on the time $t$ since formation, or on the mean age; {\it $u-v$ is expected to be a constant for all galaxies undergoing secular evolution, irrespective of age!} Instead $u-v$ responds distinctly to quenching and bursting events and is an indicator of recent change in SFR {\refbf irrespective of dust extinction}. 
This also implies that a $u-g$ index will have no more information on the age of a galaxy than $v-g$ and instead only more noise due to the fainter $u$ fluxes of galaxies. SDSS, in contrast, has a single, broader $u$ band, so that the single diagram of $g-i$ vs. $u-g$ contains all information on the SFH of a galaxy that is discernible from SDSS colours alone. 

Pending an investigation of dust effects, it is conceivable that the SkyMapper filter set can disentangle the effects of mean age of the population and recent changes in the star formation history, while SDSS passbands offer too few constraints to disentangle these two parameters from $ugriz$ alone. While this is a promising outlook for further exploration in a separate paper, what matters for this paper is the space covered by galaxies relative to AGN.

\subsection{The QSO locus expected in SkyMapper colours}

Figure~\ref{CC} also shows the colour locus for QSOs and nuclei of bright Seyfert~1 galaxies at redshift 0. We take the QSO spectral template from \citet{VandenBerk01}, separate the emission-line contour from the continuum and reassemble them into a grid of QSO properties. We span a 2D-space of QSO SEDs using a range of slopes multiplied onto the mean continuum (from $\Delta \alpha = +1$ to $-1$), and independently vary the intensity of the emission-line spectrum by factors from $0.33\times$ to $2.4\times$ relative to the mean. 

There is little distinction between the QSO and galaxy loci in $griz$. This may come as a surprise given the usually quite blue power law slopes of QSO continua. However, the QSO template used here was built from empirical spectra in the wavelength range of the SDSS spectrograph. The optical section of the template is thus dominated by contributions from low-redshift QSOs, which are largely of lower luminosity and thus show contributions from their host galaxies. 
{\refbf
While the host galaxy contribution explains the break in the power law of the template continuum near 400~nm, as shown convincingly by the improved QSO template of \citet{Selsing16}, the SDSS-based mixed template is appropriate for the low-redshift AGN in this work.}

On the blue side of 400~nm, however, QSO and galaxy SEDs deviate from one another, and the $u-v$ index may provide the clearest distinction of any single measure. Naturally, we assume that the nuclear SED of an AGN can be approximated by a blend of light from a stellar population and an active nucleus. Seyfert-2 galaxies mostly add narrow emission lines and optionally a weak featureless continuum, which have never been successfully differentiated in broad-band optical photometry from the SEDs of star-forming galaxies. Seyfert-1 galaxies of increasing nuclear luminosity will form a mixing sequence from their nuclear stellar population to their relevant QSO SED. A similar mixing sequence arises when using SEDs from apertures of different sizes that include different amounts of host galaxy light.

We note that \citet{Pol17} provide a spectral template specifically for Seyfert-1 galaxies, whose spectral index is redder at wavelengths of $\lambda>400$~nm by up to $\sim 0.5$ at the lowest redshift. However, this difference results from an increased contribution from the host galaxy as evidenced also by an increased strength of stellar absorption lines in the template. We continue to start from the somewhat purer QSO SED of \citet{VandenBerk01} and explore mixing of QSO+galaxy SEDs in Sect.~\ref{Results}. 

Finally, we investigate briefly the utility of adding the GALEX near-ultraviolet (NUV) band to the SEDs. We find the $NUV-u$ colour index to range from $\sim -0.5$ in starbursts to $\sim +5.5$ in old populations, while the QSO grid at redshift 0 ranges from $\sim -0.2$ to $\sim +0.7$. Because of this overlap in colour between AGN and starbursts, the GALEX NUV band does not add any discriminating power to the SkyMapper $u-v$ index. 

\subsection{The redshift dependence of the SkyMapper $u-v$ index}

Given the intriguing possibility that SkyMapper $u-v$ could be used as a burst-and-quench or AGN indicator, we investigate how it changes with redshift. As galaxy SEDs are redshifted, their A star features will get increasingly out of sync with the two passbands $u$ and $v$, so the $u-v$ index will lose its diagnostic power as a burst-and-quench indicator eventually. 

We recalculate the colour locus in steps of $\Delta z = 0.01$ and find an increasing dispersion of event-free colours. This is, because the two filters no longer straddle the 365~nm Balmer break; at $z=0.1$ both the $u$ and the $v$ filter probe entirely the rest-frame near-UV region below 365~nm, and hence the $u-v$ index shows simply a UV slope that depends on star formation and mean age of the stellar population. We also find a trend whereby the mean $u-v$ colour of normal galaxies gets bluer with redshift, reducing the difference between normal galaxies and AGN. These effects tend to diminish the power of the $u-v$ index as an AGN indicator at redshifts of $z>0.05$. While redshift is known a-priori for many galaxies and its effect can be calibrated, the decline in discriminating power can not be overcome. We revisit this point with empirical data in Section~\ref{data}.

\section{Data from S{\sevensize\bf ky}M{\sevensize\bf apper} DR3}\label{data}

The third data release (DR3) of the SkyMapper Southern Survey differs from the previous DR2 \citep{Onken19} only in data volume and sky coverage, while the nature of the data products is the same. DR3 provides full six-band coverage from Main-Survey exposures across three quarters of the Southern hemisphere, which is more than twice the fraction of DR2. However, not all of the planned visits to the relevant sky tiles are finished, and due to gaps in the CCD mosaic parts of many large galaxies are still affected by incomplete or shallower data. DR3 provides images in native seeing as well as catalogues with photometry obtained on single-visit images. While images can be PSF-homogenised and co-added in principle for the purpose of creating galaxy colour maps, we will not need this here\footnote{Large galaxies in DR3 images are affected by an issue, which masquerades as a background over-subtraction but really stems from the bias-removal procedure in the data reduction: the CCD readout electronics of SkyMapper are unstable on timescales that are shorter than the read-out time of a single CCD row. Hence, the CCD overscan does not inform the bias level and its variation across a given row of an image. Instead, the shape of the bias is determined by an empirical PCA-based process after masking detectable objects from a frame \citep{Wolf18}. Faint outer regions of a galaxy within read-noise level will not be masked but affect the bias fit and are thus partially subtracted. This affects only the $u$ and $v$ band images, because these are (always) read-noise limited. The cores of bright galaxies and most point sources, in contrast, are expected to be affected very little. 
}. 

In this paper, we use the catalogue photometry provided in the release and investigate only central colours of galaxies, which are also most relevant to the detection of active nuclei.
As we focus on low-redshift galaxies, we use the large samples of galaxies with known redshift from the nearly hemispheric 6dF Galaxy Survey \citep[6dFGS;][]{Jones04} and from the all-sky 2MASS Redshift Survey \citep[2MRS;][]{Huchra12}.

\subsection{Nearby galaxy sample in DR3}

6dFGS contains $\sim 114,601$ sources with recession velocity $cz>300$~km\,s$^{-1}$ and redshift quality flags of 3 or 4, {\refbf which means that over 95\% of the redshifts are considered reliable}. Over 98\% of them have a counterpart in SkyMapper within a matching radius of $2\arcsec$. We further require them to have in each of the six bands at least one Main Survey image with a photometric measurement of good quality (\texttt{FLAGS}$<4$ and \texttt{NIMAFLAGS}$<5$). The shallower 2MRS contains 25,398 sources with $v>300$~km\,s$^{-1}$ and a counterpart in DR3 within $2\arcsec$, most of which are duplicate with 6dFGS. The additional objects include very bright galaxies and galaxies at low Galactic latitude outside the 6dFGS footprint. We complement the selected 6dFGS dataset with missing sources from 2MRS, and given that those have larger footprints in the images, we include measurements with \texttt{NIMAFLAGS}$<20$, whose central photometry is still expected to be reliable.

DR3 contains tables of the galaxy catalogues that are already cross-matched to the SkyMapper \texttt{master} table, which can then be further joined with the \texttt{photometry} table using the unique \texttt{object\_id}. The \texttt{photometry} table contains one row for every single detection of an object in a unique image and can be joined with the \texttt{images} table using the \texttt{image\_id}. The entry \texttt{exp\_time} in the image table reveals the type of image, where Main Survey images all have 100~sec, while Shallow Survey images range from 5 to 40~sec depending on the filter. 

With a suitable SQL query, we selected objects with Main Survey photometry in all six bands and averaged their magnitudes among the repeat visits within any filter. This results in a sample of 25,958 galaxies in 6dFGS, complemented with 1,527 galaxies from 2MRS, with deep SkyMapper $uvgriz$ photometry.

\begin{figure}
\begin{center}
\includegraphics[width=\columnwidth]{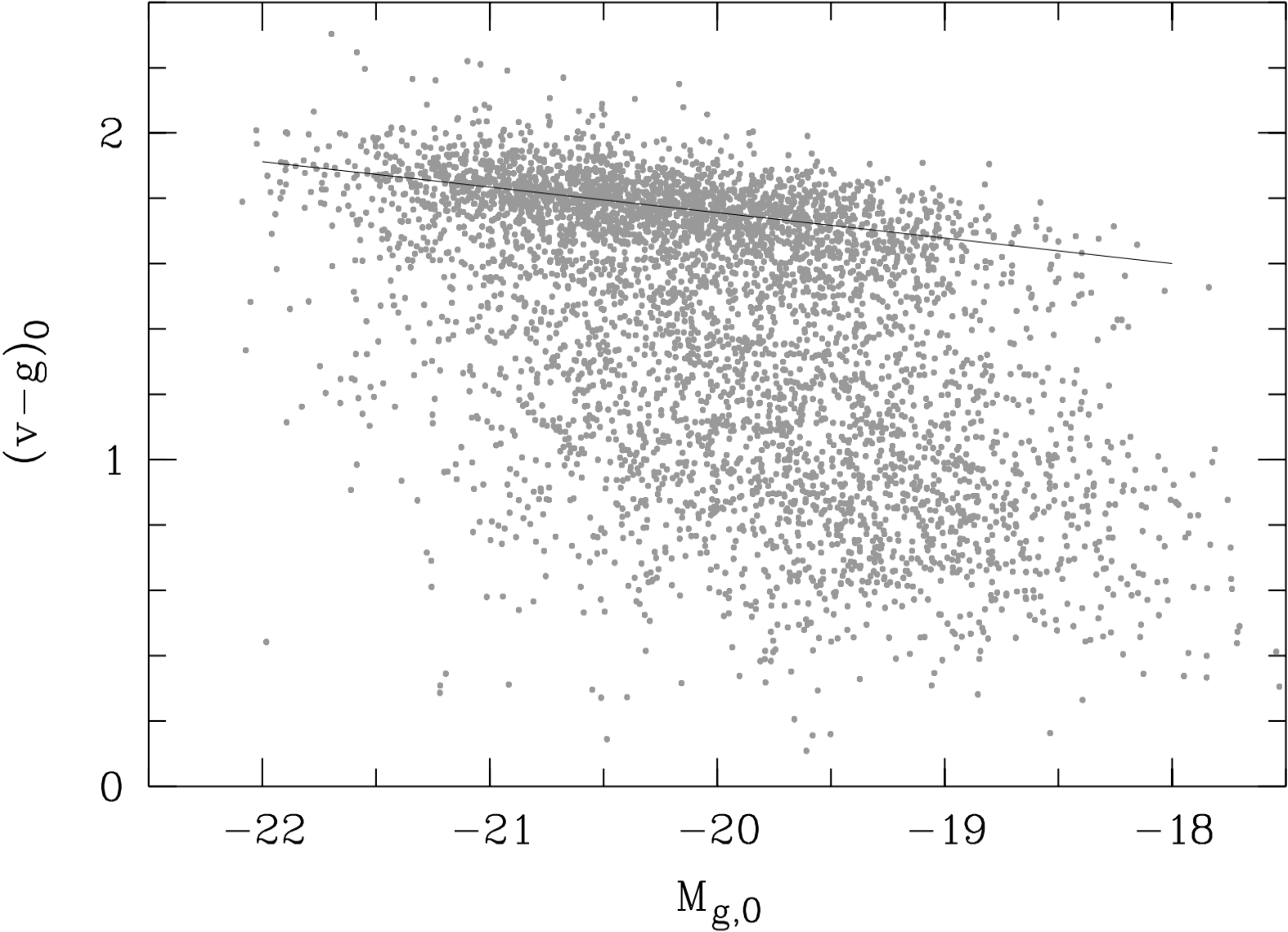}
\caption{Colour-magnitude diagram of 6dFGS galaxies in the redshift range of $4,000 < v < 8,000$~km\,s$^{-1}$, using $5\arcsec$-aperture colours from the DR3 catalogue and Petrosian $g$ band luminosity. The line is a standard colour-magnitude relation fit to red-sequence galaxies.
}\label{CMD}
\end{center}
\end{figure}

\subsection{Properties of DR3 galaxy photometry}

The \texttt{photometry} table contains a sequence of  magnitudes in apertures ranging from $2\arcsec$ to $30\arcsec$ diameter, as well as Petrosian and PSF magnitudes. The aperture magnitudes are corrected for aperture losses assuming a PSF shape (while the aperture flux columns are left uncorrected to preserve the original measurements). Seeing-induced aperture losses are thus compensated and aperture magnitudes will vary less with seeing changes than is usually the case for extended sources. They are, however, not perfectly stable against seeing variations since the varying amount of light scattered into the aperture from outside is not compensated \citep[for more detail see][]{Onken19} . 

We first use the 6dFGS sample to investigate the usefulness of the DR3 catalogue photometry in describing nuclear photometry of nearby galaxies. The fraction of total galaxy light captured by the apertures, and hence the contrast of the nucleus, depends obviously on the distance of a galaxy and on its physical extent. 

To this end, we inspect the colour scatter in the red sequence of predominantly passive galaxies. We use galaxies in the redshift range of $4,000 < v < 8,000$~km\,s$^{-1}$ and exclude faint galaxies with a $5\arcsec$-aperture magnitude of $g_{\rm APC05}>16$. 

We de-redden the photometry in $uvg$ bands with dust maps from \citet{SFD98} and filter coefficients derived in \citet{Wolf18} using a \citet{F99} extinction law. From the Petrosian $g$ band magnitude \texttt{g\_petro} we determine a luminosity $M_g$ (without K correction given the low redshift of 0.02). 
After selecting red galaxies we obtain linear fits for the colour-magnitude relation, using $5\arcsec$ aperture colours, and find:
\begin{eqnarray}
   &    (v-g)_0 = 1.756 - 0.078(M_{g,0}+20)   \\
   &    (u-v)_0 = 0.397 - 0.003(M_{g,0}+20)
\end{eqnarray}
The absence of a slope in the $u-v$ colour-magnitude relation confirms our expectation that the $u$ band adds no sensitivity to age or metallicity and the $v-g$ index is a sufficient proxy for $D_{4000}$.

With an rms of 0.075~mag in $(v-g)_0$ and 0.05~mag in $(u-v)_0$, this aperture has the smallest colour scatter in the red sequence. Towards larger apertures the scatter increases, most likely due to colour gradients in early spirals mixed into the red sequence. In the smaller $3\arcsec$ aperture, the rms scatter also increases, probably due to noise and seeing variations, to 0.115~mag and 0.068~mag respectively. We thus continue to work with $5\arcsec$ aperture colours.

\begin{figure}
\begin{center}
\includegraphics[width=\columnwidth]{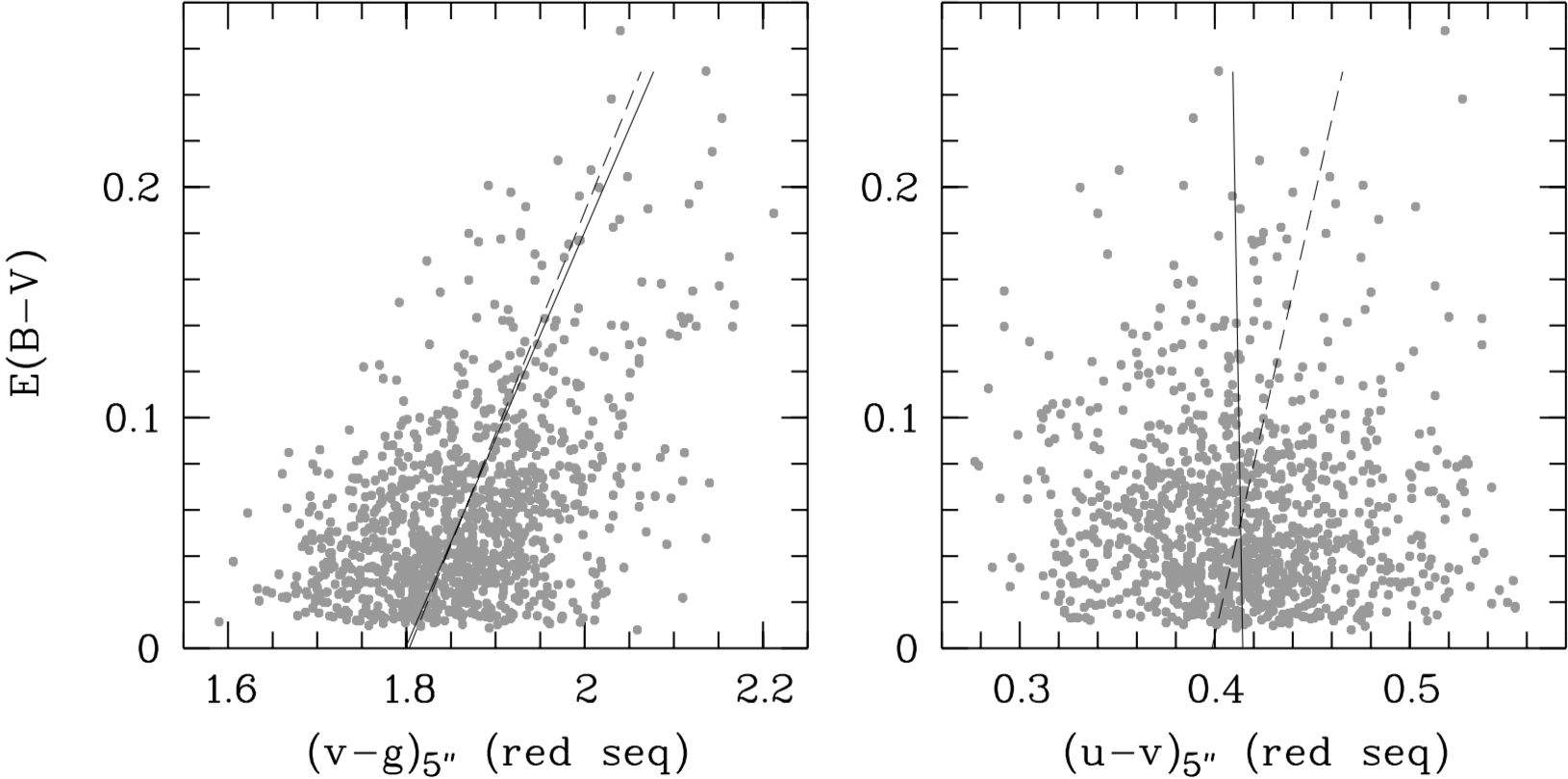}
\caption{Colours of red-sequence galaxies vs. foreground reddening. Solid lines are fits to the data, dashed lines use reddening coefficients from \citet{Wolf18}. The empirical fit confirms the redddening coefficient for $v-g$ but suggests that no dereddening is required for $u-v$. }\label{dust}
\end{center}
\end{figure}

\subsection{Reddening correction}

We test the dust correction empirically by looking for correlations in nuclear galaxy colour with interstellar foreground reddening, which ranges from $E(B-V)\approx 0.01$ to $0.27$ in this sample. We restrict the comparison to red-sequence galaxies, which have been successfully used as standard crayons by \citet{PG10} given their tight and homogeneous colour distribution. 

\citet{Wolf18} derived synthetic reddening coefficients from the SkyMapper filter curves and the \citet{F99} reddening law, finding $R_v-R_g=4.026-2.986=1.040$ and $R_u-R_v=4.294-4.026=0.268$; the latter value is close to zero, because $u$/$v$ is a close pair of relatively narrow filters\footnote{According to survey documentation at \url{http://skymapper.anu.edu.au/filter-transformations}, these $R$ values are to be used with $A=R\times E(B-V)_{\rm SFD}$ and not with $A=R\times 0.86 E(B-V)_{\rm SFD}$, a misunderstanding that has affected literature on SkyMapper photometry.}. From the red-sequence galaxy sample selected in the previous section, we find that the $v-g$ correction appears appropriate. A fit to the empirical colours yields a reddening coefficient of $R^{\rm obs}_{v-g}=1.114\pm 0.065$, which is within $\sim 1\sigma$ of $R^{\rm W18}_{v-g}=1.040$. However, we find with $4\sigma$ confidence that the $u-v$ colour is over-corrected, and indeed no evidence that a dust correction in $u-v$ is required at all. The empirical fit yields $R^{\rm obs}_{u-v}=-0.020\pm 0.038$, which is within $0.5\sigma$ of no reddening and inconsistent with $R^{\rm W18}_{u-v}=0.268$. 

We note that $R^{\rm W18}_{u-v}=0.268$ is expected to reduce to $R^{\rm W18}_{u-v}=0.22$ for observations made at an increased airmass of 2, where the atmospheric extinction effectively reddens the $u$ band filter curve, and a \citet{Cardelli89} law reduces it further to $R^{\rm C89}_{u-v}=0.16$.  These are subtle and interesting facts to consider for high-quality calibrations in the $u$ band, maybe also in light of a future refinement of the filter efficiency curve; we note, that we have no knowledge of the CCD sensitivity at $\lambda < 350$~nm, i.e. for half of the filter transmission region. There is also the possibility that the photometric calibration in DR3 has a reddening-dependent bias as it is tied to Gaia~DR2 in a way that involves a reddening-dependent extrapolation \citep[see details in][]{Onken19}.

Because of these subtleties and the fact that dereddening or not makes less than a $\pm 0.02$~mag difference for $>95$\% of the sample, we choose to ignore reddening in $u-v$ and set $R_u-R_v\equiv 0$ from now on. Since $u-v$ is a burst-and-quench indicator and most galaxies are expected not to show much signal, we also remove the red-sequence cut now and find that galaxies overall have $\langle u-v \rangle = 0.410$ with an rms scatter of $0.058$~mag.

\subsection{Spectroscopic object types}

In this work, we {\refbf study the colours of galaxies with AGN and specifically investigate whether we can differentiate Seyfert-1 galaxies from other galaxies by colour. The requirement for spectra in this work is then to reveal the broad emission lines of true type-1 AGN. Spectra also allow some differentiation of type-2 AGN from non-AGN galaxies, depending on spatial resolution, but this will be a non-critical by-product}. While the 6dF catalogue specifies redshifts, it contains no information on the galaxy type. We first consider known AGN from three, partially overlapping, literature sources: 
\begin{enumerate}
    \item The Hamburg-ESO QSO Survey \citep[HESQSO;][]{Wisotzki00} covers less than 20\% of the Southern hemisphere, but is arguably the most complete survey for broad-line AGN as it is based on objective-prism spectra for AGN selection. At low redshift, this approach requires that objects display a broad H$\beta$ line.
    
    \item The RASS-6dFGS catalogue \citep{Mahony10} covers the whole 6dFGS footprint with a sample of X-ray-selected AGN in the ROSAT Bright Source Catalogue \citep{Voges99}. Its completeness ought to be very high and unbiased between narrow-line and broad-line types for objects of sufficient X-ray luminosity. Especially, obscured type-2 AGN would be included, while dormant AGN with low accretion rate are expected to be missed. However, as we will see below, quite a few Seyfert galaxies of interest for this work are not contained in this catalogue. The RASS-6dFGS table contains a classification into narrow-line and broad-line objects based on the spectral atlas of the 6dFGS.
    
    \item The Siding Spring Southern Seyfert Spectroscopic Snapshot Survey  \citep[S7;][]{Dopita15, Thomas17} has recently re-observed 131 previously known AGN with the integral-field unit WiFeS at the ANU 2.3m telescope \citep{Dopita07}. The S7 lists a finer classification into subtypes based on their spectra and is limited in redshift to $z\la 0.02$. 
\end{enumerate}
 
The combined sample is not a volume-limited complete AGN sample, but it contains AGN of all subtypes and a complete subset of optically cross-identified, flux-limited X-ray sources.

We find that the colours of this AGN sample are generally as expected: AGN with broad emission lines are the bluest objects in $u-v$, with colours ranging from $-0.2$ to $+0.15$; while those with only narrow lines can be as red as the general galaxy population, up to $u-v\approx +0.5$. Searching the complete 6dF sample for AGN by eyeballing $\sim 25,000$ spectra is beyond the scope of this paper. An alternative approach to finding AGN would be to use automated line-fitting and a classification based on line ratios. Even though 6dFGS has been around for over a decade, such a catalogue has never been published, and we are not endeavouring to do so as part of this work either. However, at $z<0.1$ we notice over 300 objects with $u-v$ colours in the range of broad-line AGN and away from the normal galaxy distribution (see Sect.~\ref{Results} for details), which we consider candidates for further broad-line AGN. 

In order to extend our AGN sample with further broad-line AGN, we visually inspect the 6dFGS spectra of our $(u-v)$-blue candidate AGN and identify 51 further type-1 to 1.8 AGN not currently documented elsewhere. For the significant fraction of 6dFGS galaxies that were not observed by the 6dFGS itself, we adopt the classification from Simbad if it specified a Seyfert type. We also inspect the 6dFGS spectra of the literature AGN sample to verify subtypes consistently. We classify the spectra by eye, using the Seyfert types of 1, 1.5, 1.8, 1.9 and 2 as defined by \citet{Osterbrock81}: Seyfert-1 galaxies feature several broad Balmer emission lines; type 1.5 denotes objects where the broad H$\beta$ line is clearly present but comparable in flux to the narrow [O{\sc iii}] lines, 1.8 denotes a barely noticeable H$\beta$ line, 1.9 an absent H$\beta$ line, and 2 a spectrum without any broad emission lines. 

\begin{figure*}
\begin{center}
\includegraphics[width=0.85\textwidth]{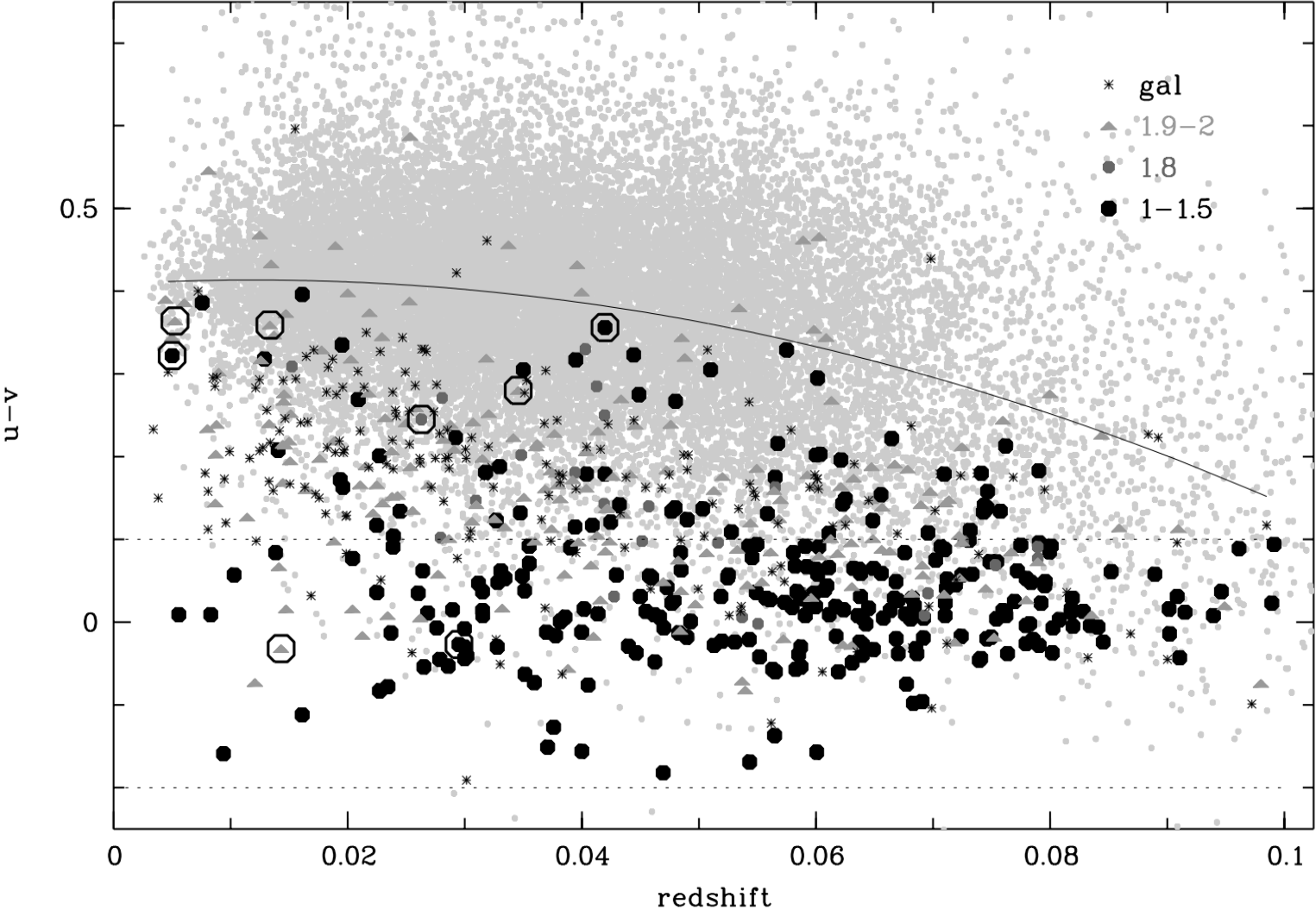}
\caption{$u-v$ index vs. redshift of 2MRS and 6dFGS galaxies: the symbols are objects typed by their spectra; circles mark known Changing-Look AGN. The solid line indicates the mean redshift trend of the overall galaxy population, and the dashed lines show the range of colours for the pure-QSO SED grid.
}\label{CzD}
\end{center}
\end{figure*}

For differentiating Seyfert-2 and star-forming galaxies we considered line flux ratios seen by eye in [N{\sc ii}]/H$\alpha$ and [O{\sc iii}]/H$\beta$ \citep[BPT diagram;][]{BPT81,Kewley01,Kauffmann03}. Here, we find several composite spectra not clearly attributed to one type. While the line ratios could be refined with automated fitting, this is beyond the scope of this paper, and the quality of composite classification has no consequence for our results. Given the spatial mixing sequence from nuclear AGN spectra to star-forming host galaxy spectra \citep{Davies14}, the AGN/composite divide depends most sensitively on spatial resolution as highlighted by the discussion of inclusion and dilution effects in \citet{Tremou15}. The spatial resolution in 6dFGS is low due to large fibres with $6\farcs7$ diameter and without better resolved spectroscopy the AGN/composite classification in this sample will remain sub-optimal. {\refbf Line-fitting and classification of the full 6dFGS spectral atlas as well as resolution issues will be tackled in a separate work to produce a quantitative community resource.}

Among the 6dFGS spectra we also notice several AGN, where a barely visible broad H$\beta$ line suggests type-1.8, while the H$\alpha$/H$\beta$ ratio appears regular (although uncertain flux calibration in 6dFGS makes this ratio unreliable); we suspect these are regular type-1 spectra diluted by a large host contribution in the 6dF fibres. This issue also relates to the discussion of low-luminosity AGN by \citet{Ho95} and highlights the problem that even a well-defined Seyfert type depends on the observing conditions.

In total, we consolidated spectral types for 688 galaxies at $z<0.1$. This includes 6 BL Lac objects, 269 AGN of type Seyfert-1 to 1.5, 23 Seyfert-1.8, 57 Seyfert-1.9 and 98 Seyfert-2 galaxies, as well as 33 spectra we interpret as Sy-2/star-forming composite spectra, 132 star-forming galaxies and 61 galaxies with absorption-line spectra, and eight known Changing-Look AGN with variable type. In the appendix we list the newly identified Seyfert galaxies that were not reported previously to our knowledge, together with their Simbad classification from May 2020.

\section{Inactive vs. active Galaxies}\label{Results}

In Sect.~2 we saw that nuclear $u-v$ colours can indicate the presence of non-obscured, type-1 AGN. The $u-v$ colour of a smoothly evolving stellar population was predicted to be $\sim 0.35$ with very narrow scatter. Colours of $|u-v-0.35|>0.03$ at redshift 0 seemed to indicate bursting or quenching signals. While starbursts seem to saturate near $u-v \approx 0.1$, the SEDs of pure unobscured QSOs are expected in the range of $[-0.2,0.1]$. Thus, a very blue nuclear $u-v$ colour indicates a strong type-1 AGN, while a more moderate blue colour implies either a weaker type-1 AGN or a strong starburst. 

In the last section, we found the mean measured $u-v$ colour to be $\sim 0.4$ at $z\approx 0.02$; this is slightly redder than predicted, which could reflect inaccurate filter transmission curves that would affect the photometric calibration of the $v$ band in particular as stellar magnitudes respond very sensitively to its filter edges, or it might be an issue with the $u$ band calibration in DR3. Despite this offset of $0\fm05$, the separation of galaxy types should work as expected.

\subsection{Redshift dependence of $u-v$ index and AGN separation}

The diagnostic power of the $u-v$ index to reveal AGN or starbursts (and quenching events) relies on the alignment of the SkyMapper filters with crucial features in galaxy spectra that is only given at low redshift. The continuum colour of AGN and starbursts is nearly independent of redshift in virtually all galaxies that SkyMapper can spatially resolve (at $z<1$), but the spectra of common normally star-forming galaxies will reveal their different Hydrogen Balmer breaks in $u-v$ only at redshifts close to 0.

Thus, we look at how the $u-v$ index depends on redshift in the bulk of the galaxy population, and compare its trend with the colours of known Seyfert galaxies. Restricting the sample to $M_{g,\rm petro}<-19$ retains more than 90\% of the galaxies at $z<0.1$, while losing only one known Seyfert-1 and four Seyfert-2 galaxies. Figure~\ref{CzD} shows the trend of $u-v$ with redshift for normal galaxies (light grey background dots), while objects with known spectroscopic types are rendered as larger symbols. Most type-1 to 1.5 AGN show bluer $u-v$ colours than the bulk of the galaxy population. We also see AGN on the normal galaxy locus, which are mostly narrow-line (type-2) AGN, whose weak featureless AGN continuum affects the $u-v$ colour very little relative to the colour of a pure stellar population.

\begin{figure*}
\begin{center}
\includegraphics[width=0.51\textwidth]{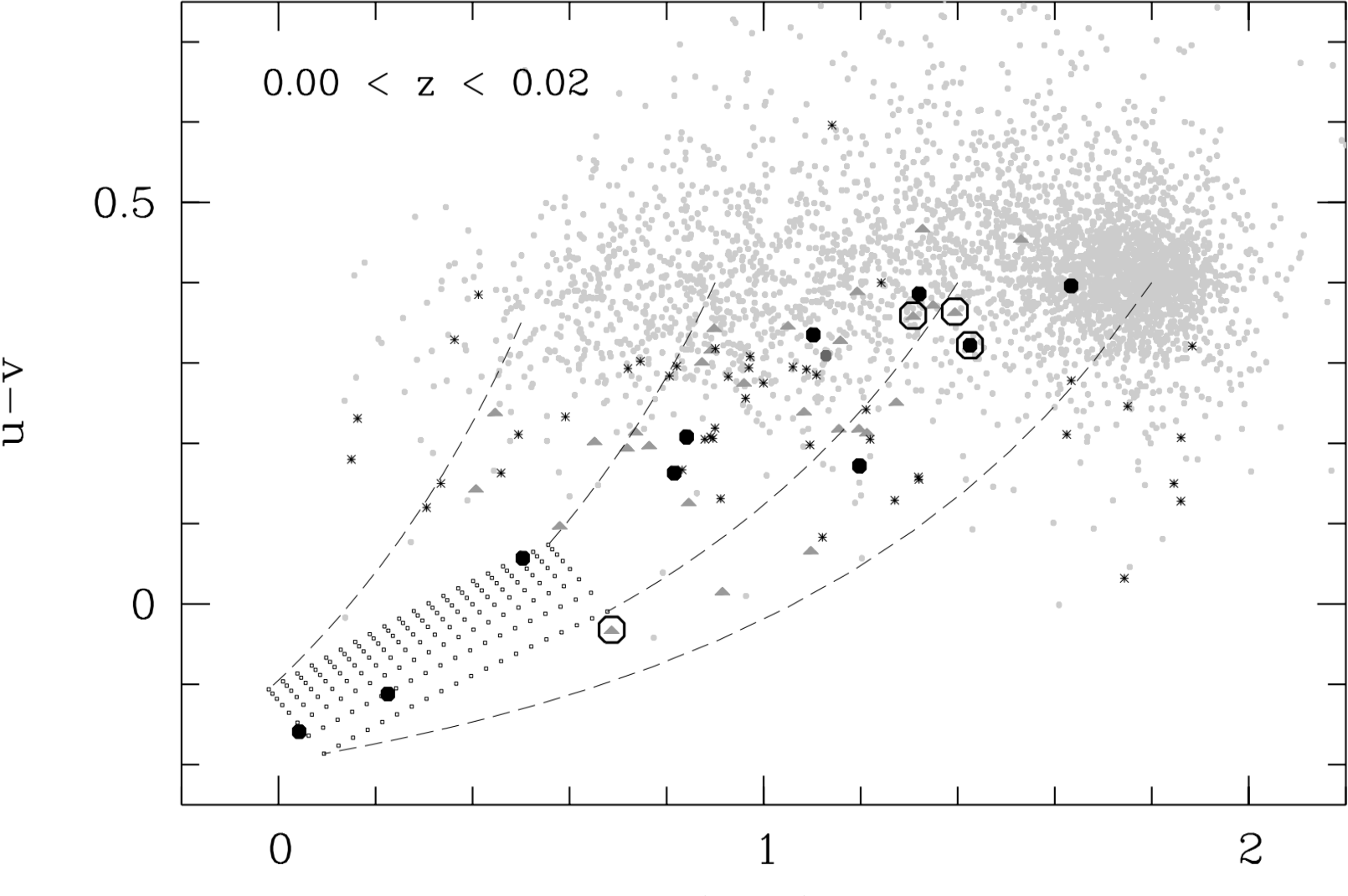}
\includegraphics[width=0.485\textwidth]{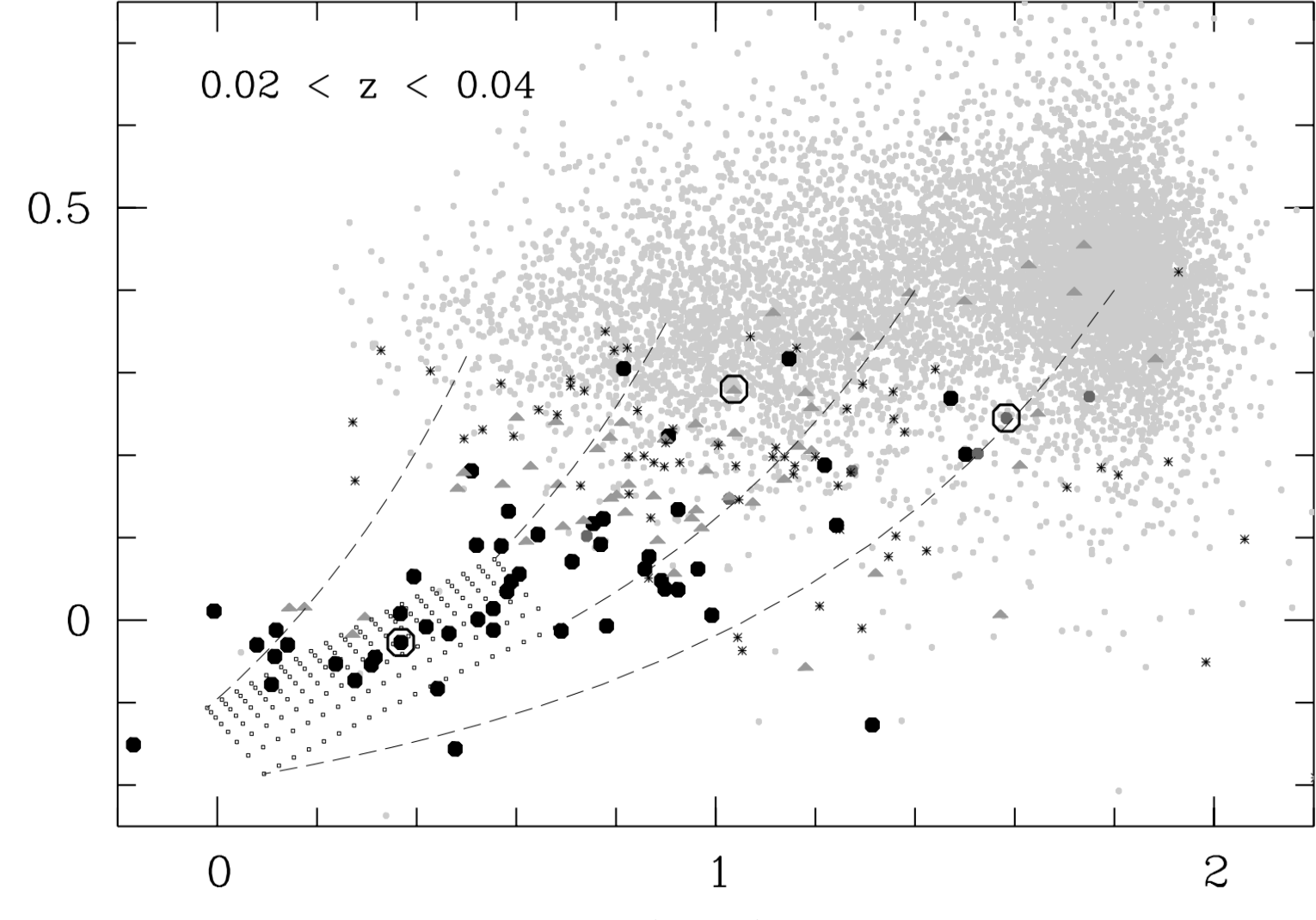}
\includegraphics[width=0.51\textwidth]{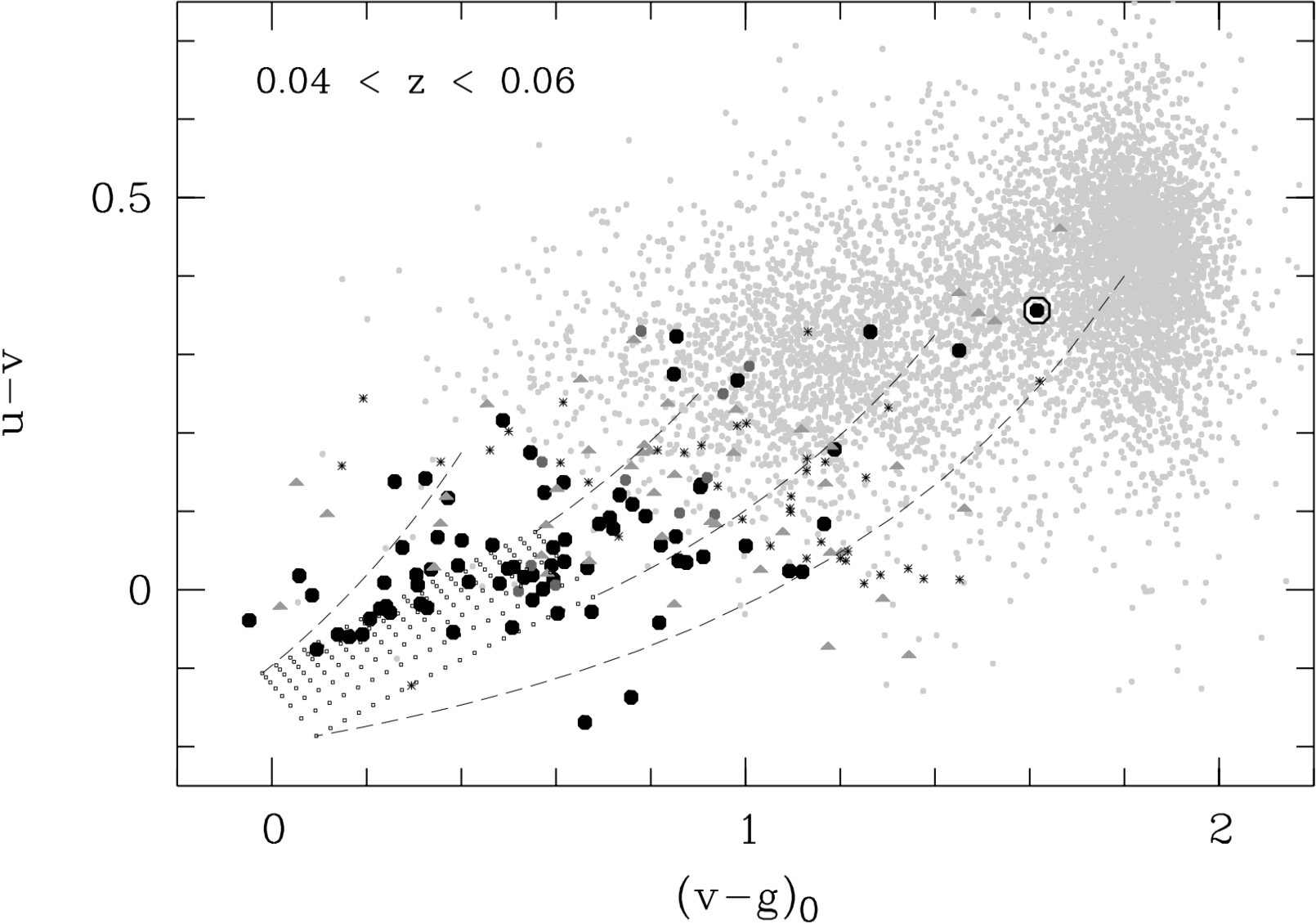}
\includegraphics[width=0.485\textwidth]{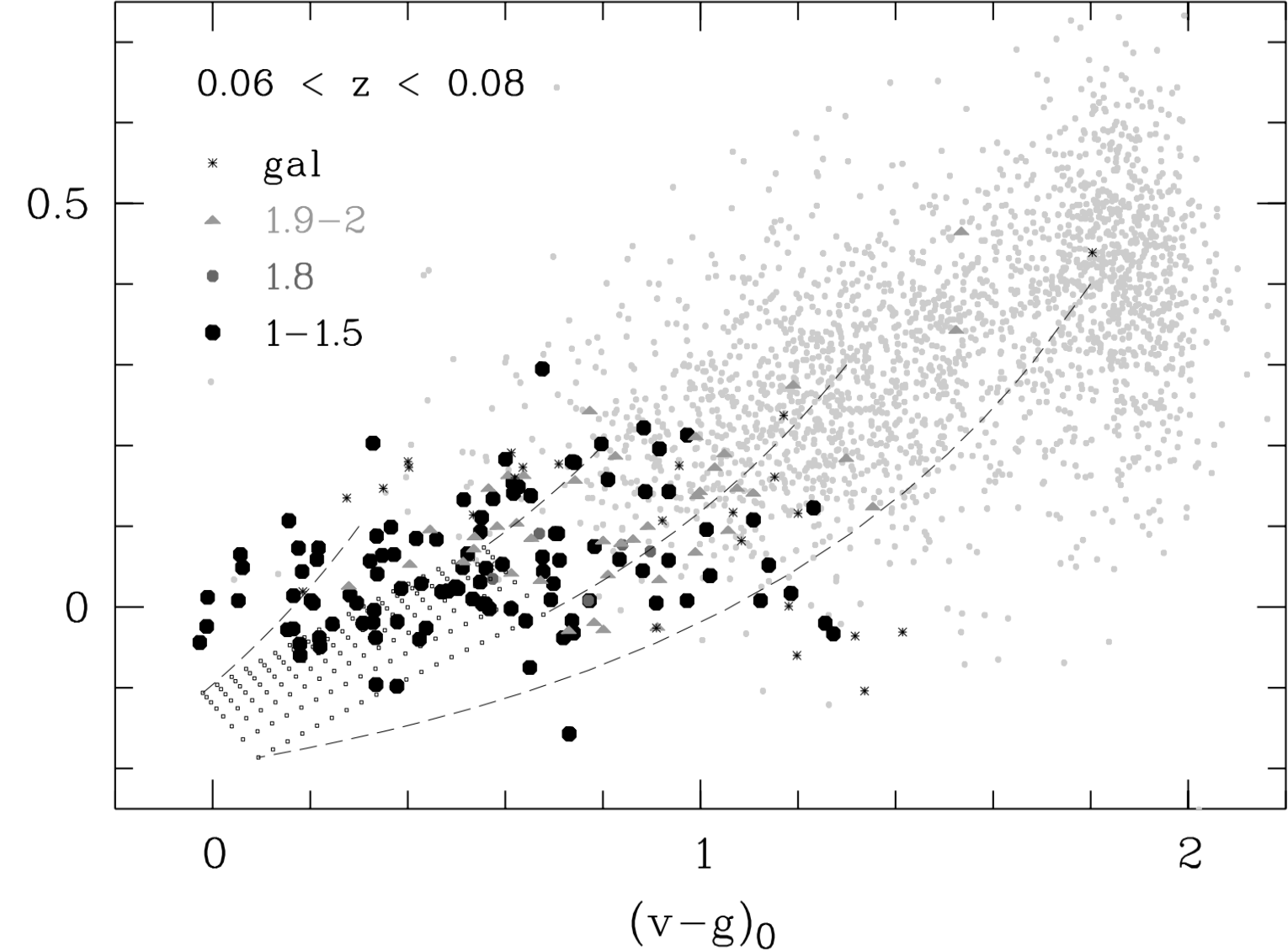}
\caption{Colour-colour diagram of 2MRS and 6dFGS galaxies: the symbols are objects typed by their spectra; circles mark known Changing-Look AGN. Dashed lines represent mixing sequences from pure galaxies to pure QSOs (grid of black dots). At redshifts close to $z=0$, galaxies with young central populations, i.e. with blue $v-g$ colours, are well separated from QSOs and Seyfert galaxies with high-luminosity type-1 AGN. With increasing redshift, these two populations overlap more, as $u-v$ measures the near-UV continuum slope instead of the 365~nm Balmer break in non-AGN galaxies. 
}\label{CCD}
\end{center}
\end{figure*}

Normal galaxies trend towards bluer $u-v$ with redshift as the Balmer break moves away from the point where it is straddled by the $u$ and $v$ band. This trend reduces the contrast between normal galaxies and type-1 AGN, whose colours are nearly independent of redshift (until the Mg{\sc ii} emission line enters the $u$-band at $z>0.2$). A second-order fit to this sample yields $\langle u-v \rangle = 0.408+0.838z+34.9z^2$. At $z\la 0.05$ type-1 AGN appear well separated from normal galaxies, while the populations blend together towards $z\approx 0.1$. At $z>0.05$ we see a concentration of Seyfert-1 AGN with $u-v \approx 0 \pm 0.1$, just as expected from the grid of QSO SED colours; these are clearly AGN with higher nuclear luminosity and subdominant host galaxy light.
Starburst galaxies lie between the QSO grid and the bulk of the galaxy population. While there is significant overlap between starburst galaxies and type-1 AGN, the colour range of $u-v<0.1$ is clearly dominated by AGN. The synthetic colours of population synthesis suggested that a young starburst that has lasted for only 10~Myr can be as blue as $u-v\approx 0.05$ at redshift 0 and $u-v\approx -0.05$ at redshift 0.05. As the starburst continues and builds up a population with an increasing age, it gets redder in $u-v$, as the number of $(u-v)$-blue O~stars remains stable while the number of $(u-v)$-red A~stars grows continuously. In practice, the nuclear photometry of a galaxy will see a mix of a pre-existing older stellar population with the recent starburst event. E.g., a new starburst with a star formation rate of 1~M$_\odot$\,yr$^{-1}$ only reaches $u-v\approx 0.1$ at $z=0$ for the first few million years when it is superimposed onto a 10~Gyr-old population with a stellar mass of $10^9$~M$_\odot$. 

Obviously, our ability to identify AGN from the $u-v$ index depends on the type and luminosity of the active nucleus as well as the central stellar population it is superimposed on, and it weakens with increasing redshift. Details can only be quantified once all the spectra in the 6dFGS atlas are classified, which might reveal further AGN at lower luminosity than have been found in this work.

\begin{figure*}
\begin{center}
\includegraphics[width=0.5\textwidth]{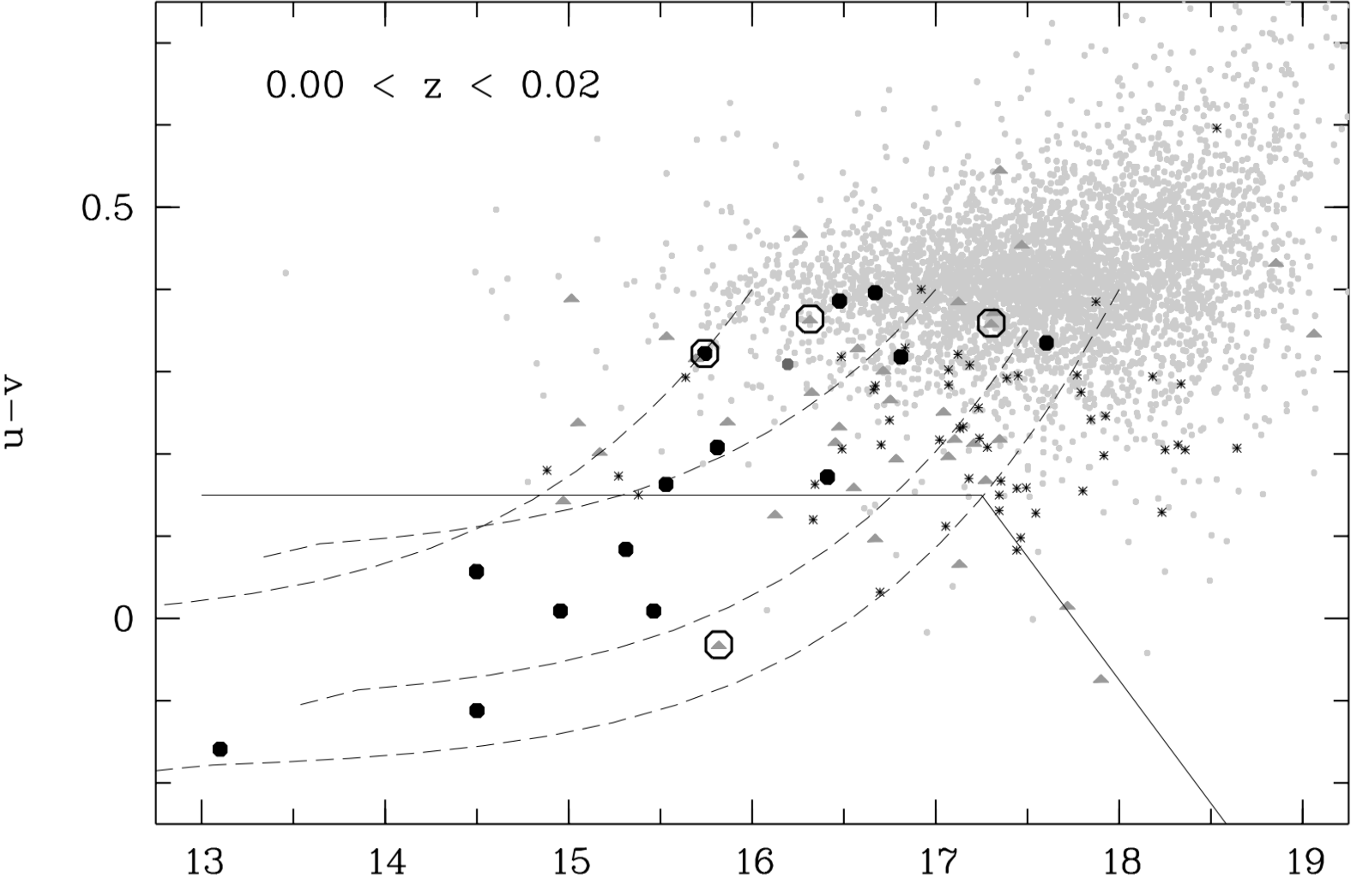} 
\includegraphics[width=0.481\textwidth]{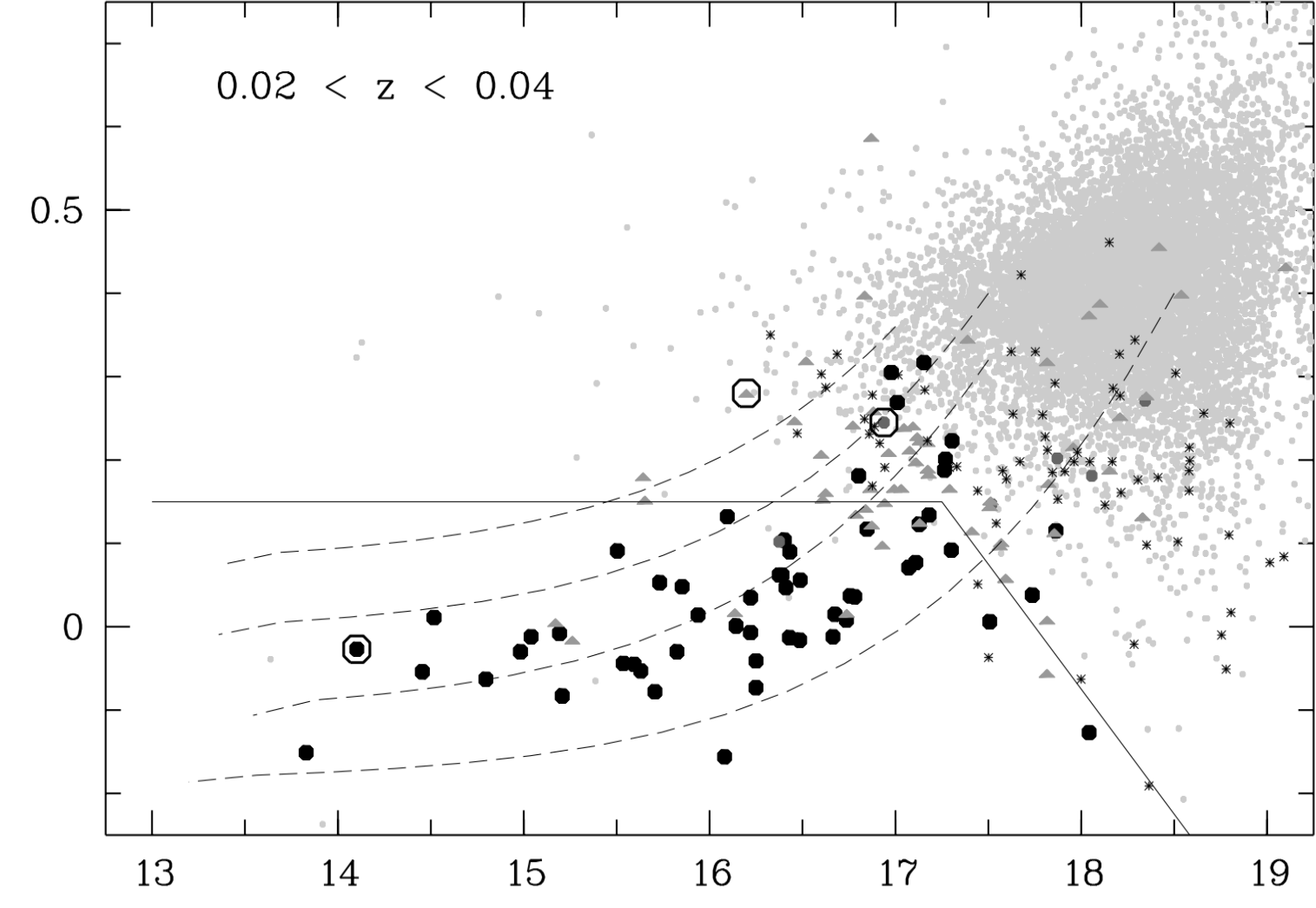}
\includegraphics[width=0.5\textwidth]{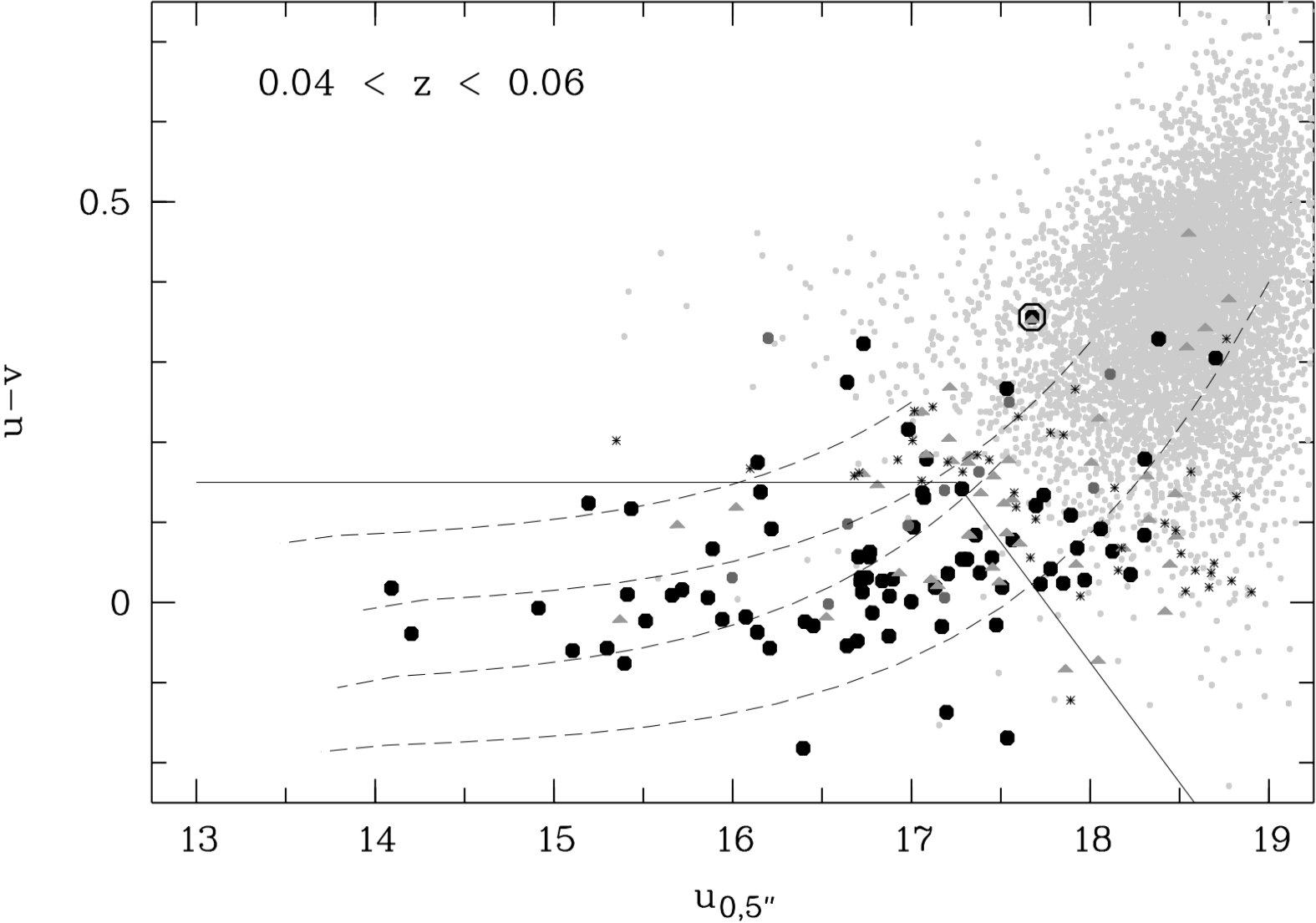} 
\includegraphics[width=0.481\textwidth]{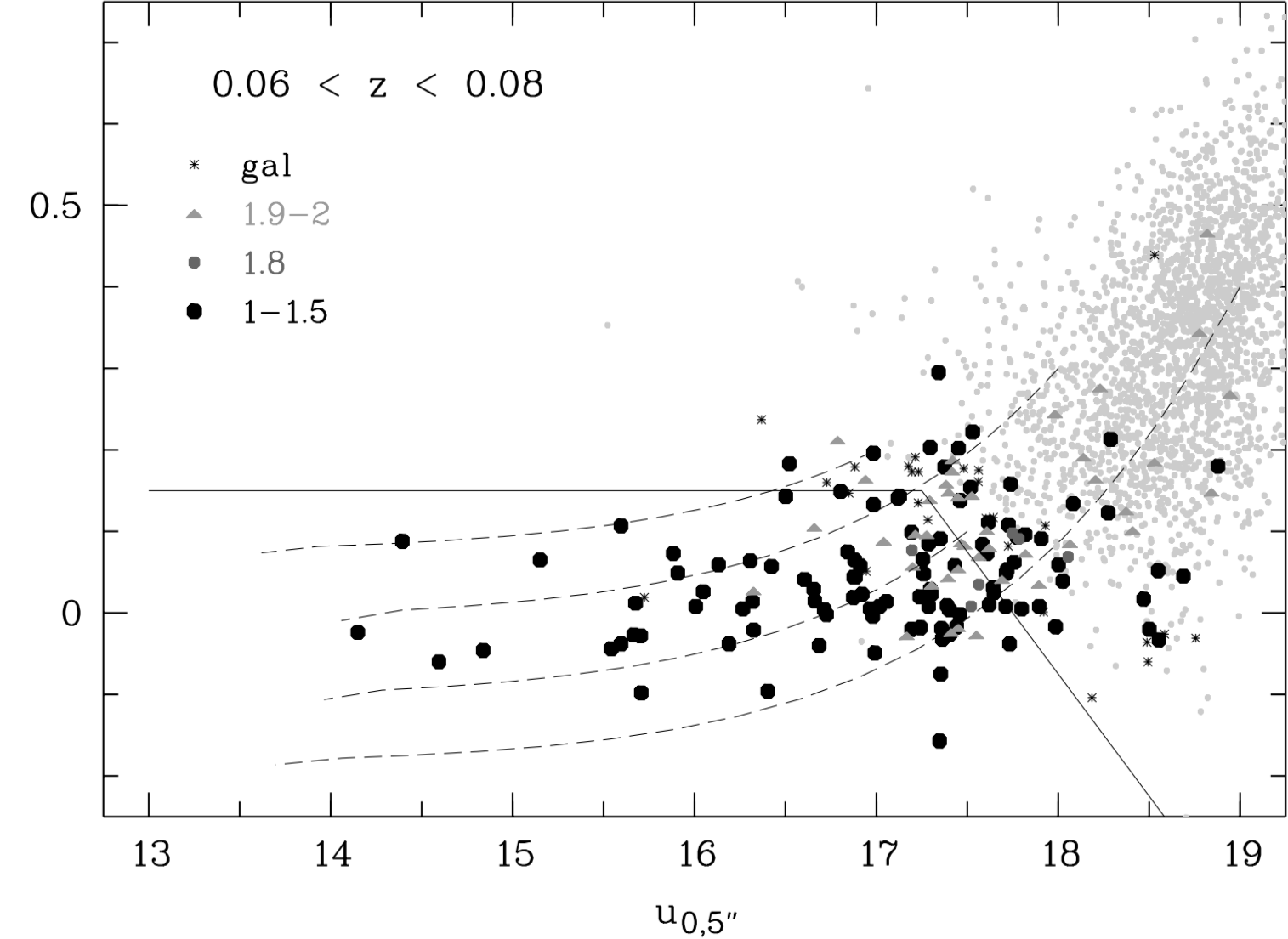}
\caption{$u-v$ index vs. $u$ band central $5\arcsec$-aperture magnitude of 2MRS and 6dFGS galaxies: the symbols are objects typed by their spectra; circles mark known Changing-Look AGN. The solid line is the selection cut for broad-line AGN from Eq.~\ref{selcut}. On average, a nuclear point-source luminosity of $M_B=-22$ corresponds to $u_{0}(5\arcsec) \approx (11,13.5,14.5,15.25)$ at $z = (0.01,0.03,0.05,0.07)$.
}\label{CLD4}
\end{center}
\end{figure*}

\subsection{A mixing sequence from star light to AGN}\label{mixing_sequence}

The SEDs of high-luminosity QSOs are dominated by AGN light, so the $u-v$ colour index clearly separates them from galaxies, but here we focus on Seyfert galaxies at lower luminosity. Their SEDs are composites of light from a stellar population and an active nucleus that form a galaxy-AGN mixing sequence. Fig.~\ref{CCD} shows the $u-v$ vs. $v-g$ colours of galaxies and known AGN. Close to $z=0$, inactive galaxies are concentrated around $u-v \approx 0.4$, as long as they are neither bursting nor undergoing quenching. Crucially, the $u-v$ colour is independent of the mean age and the absolute level of the current star-formation rate, which are well-known to strongly affect the strength of the 4000\AA -break and the $v-g$ colour, but {\it not} the 365~nm Balmer break and the $u-v$ colour. In the $u-v$ vs. $v-g$ diagram, red-sequence galaxies are centred on $(u-v,v-g)_0\approx (0.4,1.8)$, while star-forming galaxies extend horizontally out to values of $(v-g)_0 \approx 0.2$. Known AGN tend to be bluer than normal galaxies in $u-v$ irrespective of their $v-g$ colour. AGN of type~1.9 to 2 are closest to the colours of normal galaxies, but already extend bluewards from the non-AGN sequence in $u-v$, suggesting the presence of either some featureless AGN continuum or of starbursts in the central $5\arcsec$ apertures of the galaxies, a feature commonly seen in type-2 AGN \citep[e.g.][]{CidFernandes01}. 
AGN of type~1.5 to 1 are furthest offset to bluer $u-v$ colours and reach all the way to pure QSO SEDs (shown as a grid of dots). 

With increasing redshift, the distribution of normal galaxies extends to bluer colours in $u-v$, because the two filters no longer straddle the 365~nm Balmer break. The $u-v$ index then probes a UV continuum slope and depends on mean age of the stellar population. Thus, it becomes highly correlated with the $v-g$ index and leads to overlaps between young galaxies and broad-line AGN.

Some AGN seem to defy the expectation that the continuum colour of a type~1.9-2 spectrum should differ little from that of normal galaxies. However, AGN might have changed their type between the early spectral epoch and the recent SkyMapper epoch. Eight such Changing-Look AGN (CLAGN) are known in the sample and marked with circles. They include e.g. NGC~2617 at $z=0.014$, which shows a type-1.9 spectrum in 6dFGS but has QSO colours in SkyMapper. CLAGN are further discussed in Sect.~\ref{CLAGN}. 

The lines in the diagram represent four mixing sequences, which start at representative points in the galaxy SED cloud and mix in increasing amounts of AGN light from different corners of the QSO SED grid. The colours of the known AGN relative to the mixing sequences emphasise that central stellar populations in AGN cover a broad range in colour, possibly with a preference for intermediate colours, which could be studied in the future.

\begin{table}
\caption{Spectroscopic types selected by criterion of Equation~\ref{selcut}.}
\label{selcuttab}      
\centering          
\begin{tabular}{lrrrr}
\hline       
Redshift    & \multicolumn{2}{c}{$z=[0,0.04]$} & \multicolumn{2}{c}{$z=[0.04,0.08]$} \\
Object type &  (\ref{selcut}) true & (\ref{selcut}) false &  (\ref{selcut}) true & (\ref{selcut}) false \\ 
\hline
BL Lacs     &   1 &  0 &   4 &   1 \\ 
Sy 1-1.5    &  50 & 16 & 124 &  56 \\ 
Sy 1.8      &   1 &  5 &   8 &   9 \\  
Sy 1.9      &   5 & 17 &  13 &  21 \\  
Sy 2        &  11 & 43 &  11 &  27 \\  
Composites  &   0 & 18 &   2 &  13 \\ 
Galaxies    &   7 &111 &   7 &  59 \\  
CL-AGN      &   2 &  5 &   0 &   1 \\ 
\hline 
Total       &  78 &215 & 170 & 186 \\ 
\hline                  
\end{tabular}
\end{table}

\subsection{Separating type-1 and type-2 AGN}

Fig.~\ref{CCD} suggests that we may be able to separate type-1 and type-2 AGN based on the $u-v$ colour, provided the type-1 nucleus is luminous enough to move the composite SED away from normal galaxies. The contrast between AGN and stellar light depends obviously not only on the luminosity of the AGN but also on the central stellar population and the angular resolution of the aperture, hence on distance. We thus investigate the mixing sequences in terms of $u-v$ colour vs. surface luminosity in the central aperture. Surface brightness is independent of distance, except for cosmological surface brightness dimming, which we ignore here due to the small range of local redshifts; in this approximation, aperture luminosity density is equal to the extinction-corrected apparent aperture magnitude $u_{0} (5\arcsec)$ apart from an offset given by the distance modulus where a 5$\arcsec$-aperture covers a 1~kpc$^2$ area. 

Our sample mostly ranges from $u_0\approx 14$ to $19$~mag, which corresponds to a $u$ band ABmag surface luminosity density of approx. $-19$ to $-14$~mag~kpc$^{-2}$. If all light came from a nuclear point-source with a QSO spectrum, this range would correspond to nuclear luminosities of $M_B\approx -19$ to $-14$ at $z=0.01$. At our far end of $z\approx 0.08$, the $u_0=[14,19]$ range corresponds to $M_B=[-23.5,-18.5]$.

The AGN contribution makes the galaxy centre not only bluer but also brighter as illustrated in Fig.~\ref{CLD4}. The dashed lines represent the same mixing sequences as in Fig.~\ref{CCD} but also assume a brightness of the stellar population as an indicative starting point. In the lowest redshift bin, the colour of the stellar populations is so constrained in $u-v$ that colour alone may be sufficient to separate type-1 AGN from non-AGN galaxies, apart from increased scatter towards the faint end due to noise. The galaxy distribution extends to high aperture brightness or surface luminosity density, because the increase in spatial resolution towards $z\rightarrow 0$ focuses the aperture on the bright galaxy core. With increasing redshift the distribution loses its bright end as the $5\arcsec$ aperture averages the galaxy light profile over an increasing physical aperture. At this point, normal galaxies with young stellar populations show bluer $u-v$ colours that overlap with those of type-1 AGN, and the strongest indicator of a Seyfert-1 galaxy is the brightness of the nuclear region.

Using both axes, we define a selection cut for likely Seyfert-1 galaxies, which removes normal galaxies and type-2 AGN:
\begin{equation}\label{selcut}
  u-v < 0.15-\max(0,0.3\times (u_{0,5\arcsec}-17.25))
\end{equation}

Whether this rule selects Sy-1 galaxies or not depends on the luminosity of their active nuclei. Starting from the an average non-starburst galaxy with $u-v=0.4$, we calculate how much AGN light is needed to move the $5\arcsec$ photometry across the selection threshold. At $z=0.01$, a median galaxy with $u\approx 17.5$ reaches the selection line after adding a nucleus of equal $u$ band brightness. A nucleus of $u= 17.5$ corresponds to a luminosity of $M_B= -15.5$, meaning that our Seyfert-1 selection reaches a factor of 1000 below the luminosity of the classic QSO-Seyfert divide. 

There are Seyfert-1 nuclei of lower luminosity known in more nearby galaxies that are seen with better spatial resolution and stronger contrast against the host stellar population \citep{Ho95}. An extreme example is the galaxy NGC~4395, which is $16\times$ nearer than $z=0.01$ and hosts a Sy-1 nucleus with $M_B\approx -10$ \citep{FS89}. At the distance of NGC~4395 (2.6~kpc) our selection threshold would be $M_B<-9.5$, consistent with the expectation that the nucleus could be discerned in ground-based and seeing-limited observations. At the other end of the range considered here, $z=0.08$, we need to add a nucleus of $u\approx 18$ on typical host galaxies to reach the selection threshold, corresponding to $M_B\approx -19.5$.

We now consider the mix of galaxy types selected by Eq.~\ref{selcut}. At $z<0.08$ and ignoring the known CLAGN, Eq.~\ref{selcut} selects $\sim 70$\% of the known Sy-1 to 1.5 AGN and BL~Lac objects. It also selects $\sim 40$\% of the Sy-1.8, $\sim 30$\% of the Sy-1.9 and $\sim 20$\% of the Sy-2 AGN. Of the known composite objects and star-forming galaxies $\sim 7$\% are selected. In terms of purity, the cut selects 246 (non-CLAGN) objects with known type, among which $\sim 75$\% are Sy-1 to 1.5 AGN or BL Lac objects, $\sim 20$\% are Sy-1.8 to 2 AGN, and $\sim 7$\% are composites or starburst galaxies.

\begin{figure}
\begin{center}
\includegraphics[angle=270,width=\columnwidth]{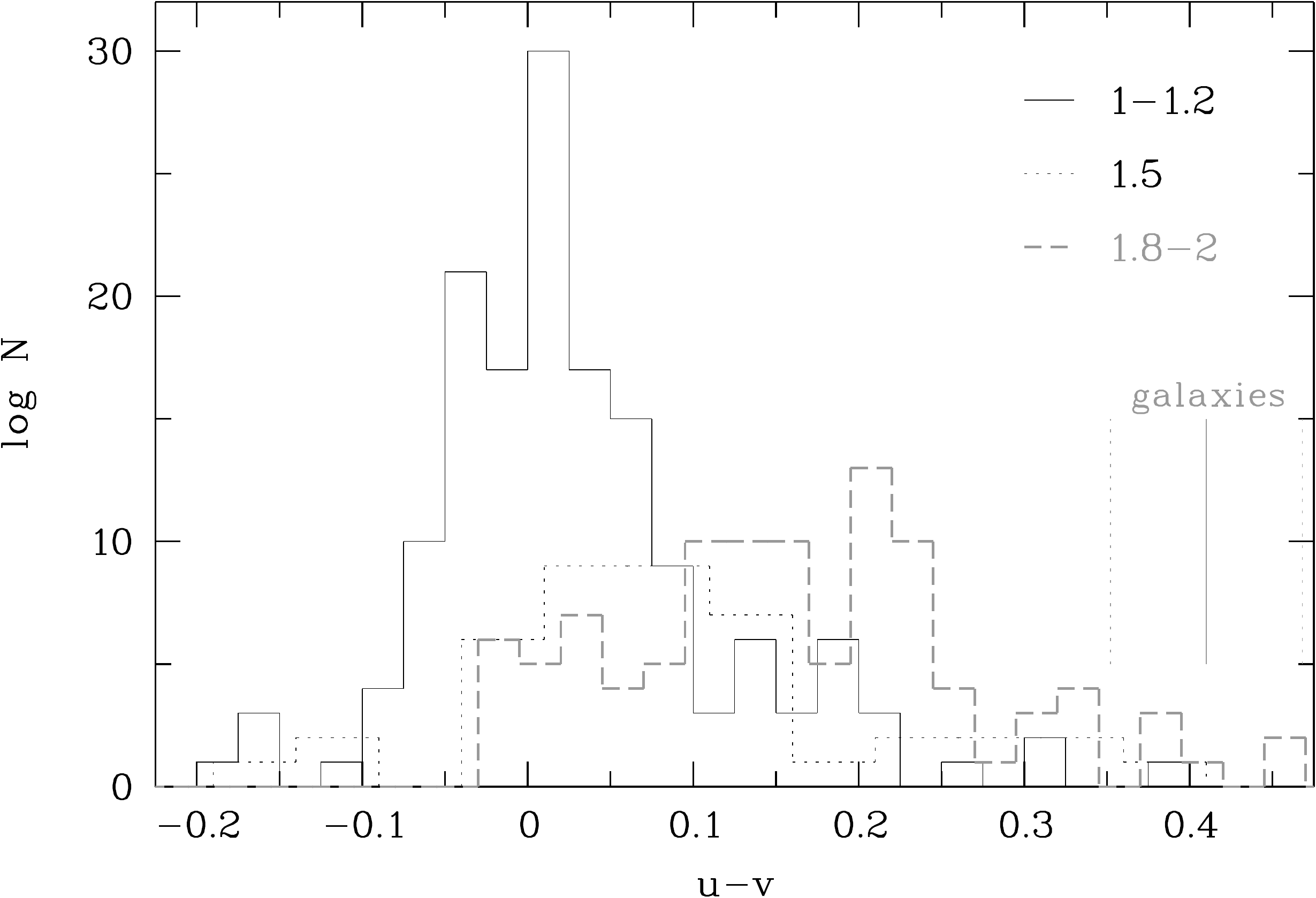} 
\caption{Histogram of $u-v$ index for different AGN types, for AGN with $u_{0}<17.5$. Normal galaxies are centred on $u-v=0.41$ with an RMS of 0.058, indicated by vertical lines.
}\label{umv_histo}
\end{center}
\end{figure}

In Fig.~\ref{umv_histo} we show histograms of $u-v$ colour for different AGN types, restricted to $u_0<17.5$. A gradual trend is evident where the Balmer breaks get redder from type~1-1.2, over type~1.5, towards type~1.8-2, although even type-2 galaxies are usually not as red in $u-v$ as normally star-forming or quiescent galaxies. There are several reasons for overlap between the types:
\begin{itemize}
    \item Intrinsic SED mixing: redder active nuclei can appear bluer when accompanied by nuclear starbursts, and bluer active nuclei will appear redder when they are only of moderate luminosity and blended with redder stellar populations.
    \item Intrinsic variability: AGN are variable sources, but the colours used in Fig.~\ref{umv_histo} are non-instantaneous and only accurate for non-variable SEDs. Even if the variability affected only brightness but not colour, the fact that different filters can be observed at different times may introduce spurious changes in colours. 
    \item Measurement errors: the formal mean 1$\sigma$-error in the $u-v$ index is $0\fm03$ and broadens the per-type distributions.
    \item Evolution: an extreme form of variability is type evolution as seen in Changing-Look AGN.
\end{itemize}

The eight known Changing-Look AGN are found on both sides of the cut. It is also possible that some of the other objects in the sample have changed their AGN type between the time their spectra were taken and the more recent photometric observations with SkyMapper. How much the overlap of AGN types in the selection diagram of Fig.~\ref{CLD4} is intrinsic and how much it is a result of objects evolving between the photometric and the spectroscopic observation cannot be answered without spectroscopy taken at the time of the photometric observations.

\begin{table*}
\caption{Known Changing-Look AGN (CLAGN) in this sample.}
\label{clagntab}      
\centering          
\begin{tabular}{lrccccl}
\hline       
Object name & SkyMapper  & redshift & 6dFGS & DR3 $u-v$  & observed & references \\ 
            & DR3 ID &          & type  & min,max  & type changes  & \\ 
\hline
NGC 7582             &   1799766 & 0.005 &   2 & $\sim +0.4$    & 2-1-1.9  & \citet{Aretxaga99,Bianchi09} \\ 
Mrk 1018             &  15202037 & 0.042 &   1 & $\sim +0.35$   & 1.9-1-1.9 & \citet{Cohen86,McElroy16} \\ 
NGC 863 =Mrk 590     &  15292183 & 0.026 & 1.8 & $+0.19$,$+0.26$& 1.5-1-2 & \citet{Denney14} \\ 
NGC 7603=Mrk 530     &  16443980 & 0.030 &   1 & $-0.10$,$+0.00$& 1-1.9-1 & \citet{TO76} \\ 
Mrk 609              &  21529338 & 0.035 &   2 & $+0.14$,$+0.38$& 1.8-1.5-2 & \citet{Rudy88,Trippe10} \\ 
NGC 1346             &  21659170 & 0.013 &   2 & $+0.25$,$+0.33$& 1-2     & \citet{Senarath19} \\ 
NGC 2617             &  75780515 & 0.014 & 1.9 & $-0.03$,$+0.08$& 1.8-1-1.8?-1? &\citet{Shappee14,Oknyansky18} \\ 
NGC 1566             & 502151340 & 0.005 &   1 & $+0.06$,$+0.42$& 1.9-1.2-1.9-1.2 & \citet{Alloin86,Oknyansky19} \\ 
\hline                  
\end{tabular}
\end{table*}

\section{Seyfert types and Changing-Look AGN}\label{CLAGN}

Several ideas have been considered when explaining the range of types in Seyfert galaxies: 
\begin{enumerate}
    \item {\it AGN orientation:} in the unification model by \citet{Antonucci93} the spectral appearance of AGN is driven by orientation. Type-1 objects provide the observer with an unobscured view of the UV-bright accretion disk and the broad emission-line region, while these components are hidden behind a parsec-scale nuclear dust torus in type-2 objects. The latter leave only the larger-scale narrow emission-line region to be seen that is caused by ionising radiation escaping from the nucleus through the open sections of the circum-nuclear dust. \citet{Ueda03} and \citet{Hasinger08} added the luminosity aspect to the unification model: AGN of lower luminosity are statistically more likely to be seen as absorbed as the opening angles of their tori may be smaller in the presence of weaker feedback.
    
    \item {\it Host galaxy orientation:} \citet{MR95} find type 1-1.5 AGN predominantly in face-on galaxies and type 1.8 + 1.9 mostly in edge-on galaxies, suggesting that the latter types are 'seen through a 100~pc-scale torus coplanar with the galactic disk\ldots and not partially obscured by an inner parsec-scale torus'. In contrast, they suggest that type-2 objects are obscured by the latter as they are seen at all orientation angles.
    
    \item {\it Evolution:} \citet{Elitzur14} proposed that the sequence from type-1 through the intermediate types to type-2 is a natural evolutionary sequence arising as the accretion rate onto a black hole declines and the broad-line region disappears from the weakening wind. In this picture, two kinds of type-2 AGN need to be distinguished: obscured type-2 AGN as proposed in (i) and (ii), and separately unobscured 'true type-2' AGN discussed already earlier by \citet{Tran01,Tran03}, \citet{Panessa02} and \citet{Laor03}. 
    
    \item {\it Changing-Look AGN:} AGN may exhibit changes in their type when accretion turns on or off. While this phenomenon is also an evolutionary process, the remarkable feature is that the drastic changes in accretion rates and the related (dis-)appearance of a broad-line region may happen on timescales of years or less. In contrast, the narrow emission-line region is so extended on a kiloparsec scale that light travel time alone delays and filters by thousands of years any response to the evolution of the ionising nuclear source. 
\end{enumerate}

Observations of CLAGN have seen them evolve between all the defined subtypes. We note here that \cite{Hawkins04} report the discovery of 'naked AGN' with strongly variable continua in long-term monitoring and narrow-line spectra. \cite{Hawkins04} argues that these AGN are in a transition stage towards a dormant black hole, because the luminosity of the objects has generally declined and the type-2 spectrum is observed only in hindsight.

\subsection{New Changing-Look candidates from SkyMapper DR3}\label{newCL}

Whether the physical difference between type-1 and type-2 is dust obscuration or accretion rate, the observable spectra are similar. While the types 1 vs. 2 are defined by the presence of broad spectral lines, their presence is linked to the visibility of a blue featureless continuum from an accretion disk. Hence, we expect that spectral type could be inferred with some certainty from photometric SEDs that probe the central continuum SED of galaxies. Objects selected by Eq.~\ref{selcut} are very likely to be type-1 AGN, while objects outside of this cut are more likely to be type-2 AGN or inactive galaxies.

The conclusion is that objects with type-1 spectra in 6dFGS that are not selected by Eq.~\ref{selcut} are good candidates for present-day type-1.8 to 2 AGN and thus may be turn-off Changing-Look AGN. Conversely, we expect that those 46 objects selected by Eq.~\ref{selcut} that have type-1.8 to 2 spectra in 6dFGS are good candidates for present-day type-1 AGN and may be turn-on Changing-Look AGN. However, these objects might well have been type-1 AGN all along as a more nucleus-focused spectrum might reveal, but in order to test for any change, modern-epoch spectra need to be extracted with $6\farcs7$-apertures for a fair comparison.

A further {\refbf 16 star-forming galaxies} appear with bright nuclei and a blue Balmer break. {\refbf We consider them contaminants of the broad-line AGN candidate list, while their origin may be either due to real blue transients such as tidal disruption events (TDE, see Sect.~\ref{TDE}), extreme starbursts, or due to data artefacts as discussed in the following section. }

\subsection{Instantaneous colours}

{\refbf
All colour indices used in this work so far are averaged over all the available data and thus not instantaneous. It is not even guaranteed that the recorded mean colours represent the true mean colour of an individual object: seeing variations affect aperture photometry of extended objects and can thus introduce mild but spurious variability into the light curve of any single passband. When colours are made from two bands, they are only expected to be precise when the individual visits in the two bands occurred under a similar mix of conditions, which is not always the case. This issue introduces additional scatter into the distilled mean colours.

However, in the SkyMapper Southern Survey, the $uv$ bands are usually observed within minutes of each other, which allows near-instantaneous colour measurements. Variability in the light curves of extended objects that results from seeing variations will cancel out in the colours defined by bands with similar seeing levels as is the case for the two bands $u$ and $v$ that are very close in wavelength. 

Variability that is coincident between two bands is often used as a strong indication of its true significance, see e.g. work on stellar flares in SkyMapper by \citet{Chang20}. For the particular case of variability in low-redshift AGN, the mixing sequences in Fig.~\ref{CLD4} also illustrate how a true change in nuclear brightness is expected to be strongly related to a change in $u-v$ colour, providing further scrutiny on the nature of any detected variability signal. This point will be illustrated in the following section, where we investigate the light curves of known CL-AGN alongside instantaneous $u-v$ colours.}

\subsection{Known CLAGN in this sample}

The SkyMapper colour index values discussed above are clipped weighted averages of measurements spread over the years 2014 to 2019, which may have averaged over low and high states and lowered the contrast for discovering changes. 
We thus discuss light curves of these objects with an emphasis on changes in the $u-v$ index. Now, we average photometry from multiple exposures only if they are taken in the same night, which usually happens only for Main Survey $uv$ exposures. The derived $u-v$ indices are thus from single nights\footnote{We allow for one exception: a $u-v$ index for Mrk~609 in late 2016 is formed from data separated by 63 days.}. 
We include photometry from Shallow Survey images\footnote{Exposures in the Shallow Survey are on average 1.0 and 1.75~mag less deep than Main Survey images in $u$ and $v$ band, respectively.} as well as long as their magnitude errors are $<0.05$.

\begin{figure}
\begin{center}
\includegraphics[width=\columnwidth]{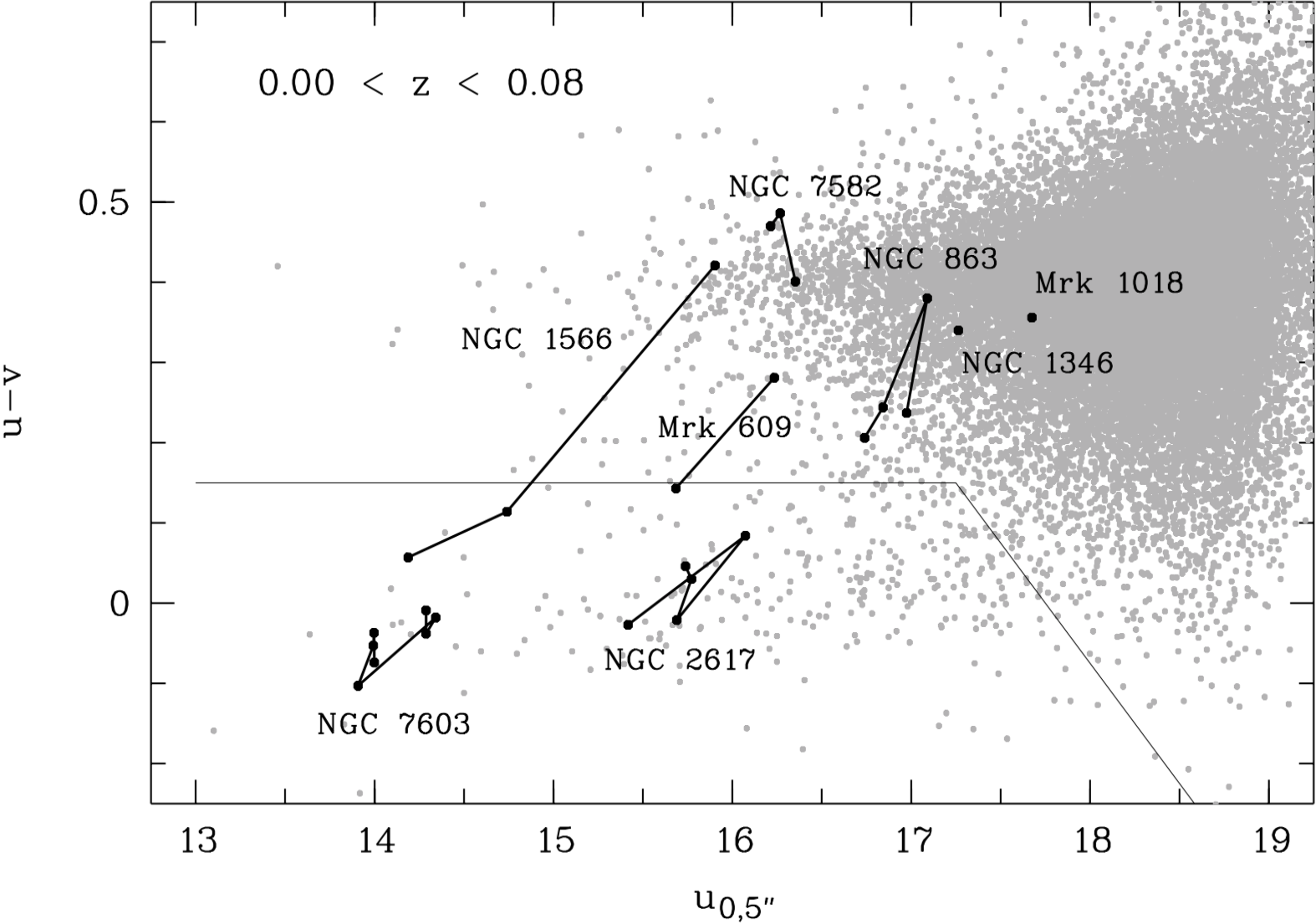} 
\caption{$u-v$ index vs. $u$ band central aperture magnitude of galaxies (grey): paths mark the evolution of known Changing-Look AGN observed in DR3. The solid line is the selection cut for broad-line AGN from Eq.~\ref{selcut}. NGC~1566 is clearly crossing the line from type-2 to type-1 during DR3.
}\label{CLD_CL}
\end{center}
\end{figure}

\begin{figure}
\begin{center}
\includegraphics[angle=270,clip=true,width=0.975\columnwidth]{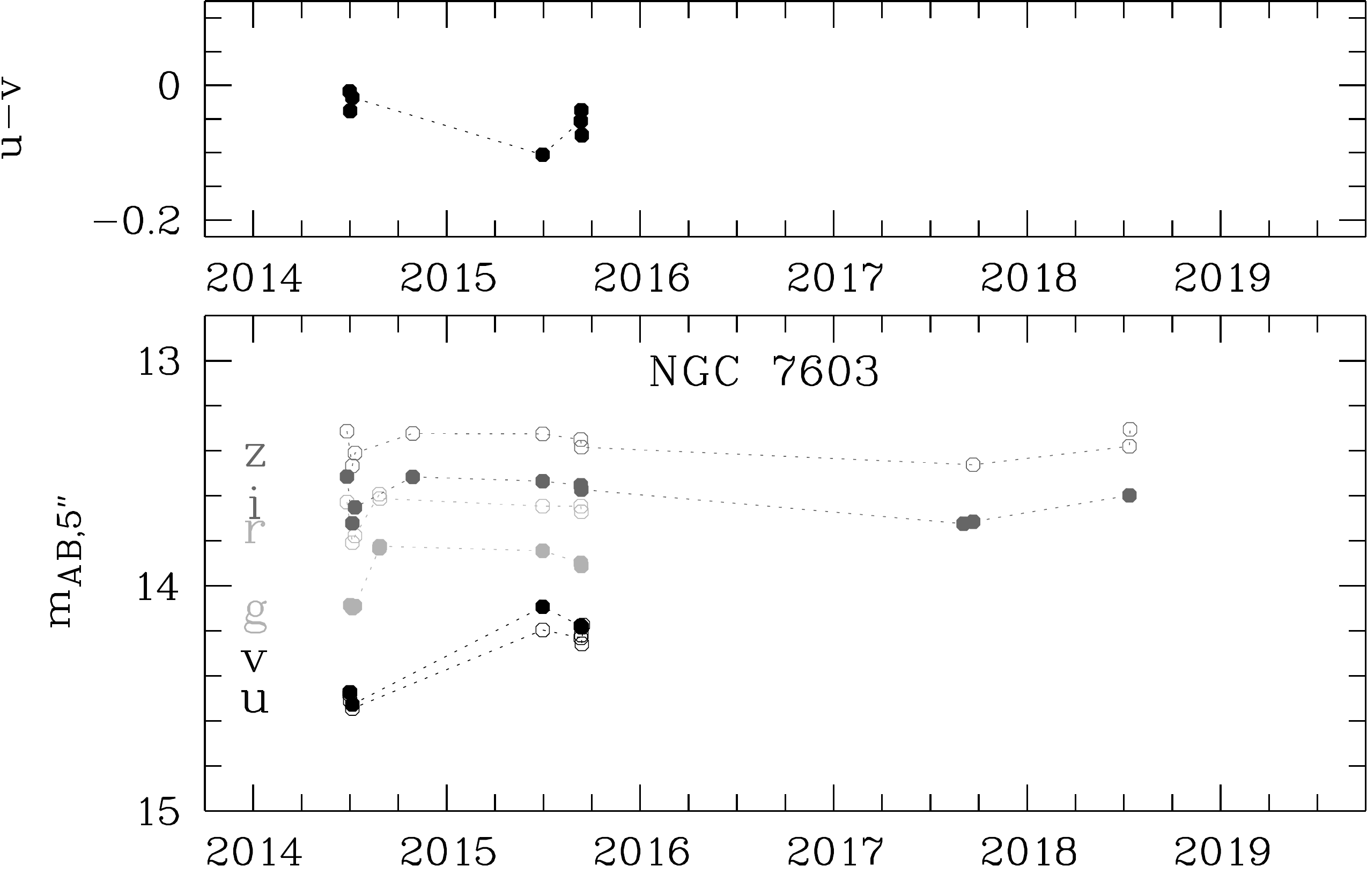} 
\includegraphics[angle=270,clip=true,width=0.975\columnwidth]{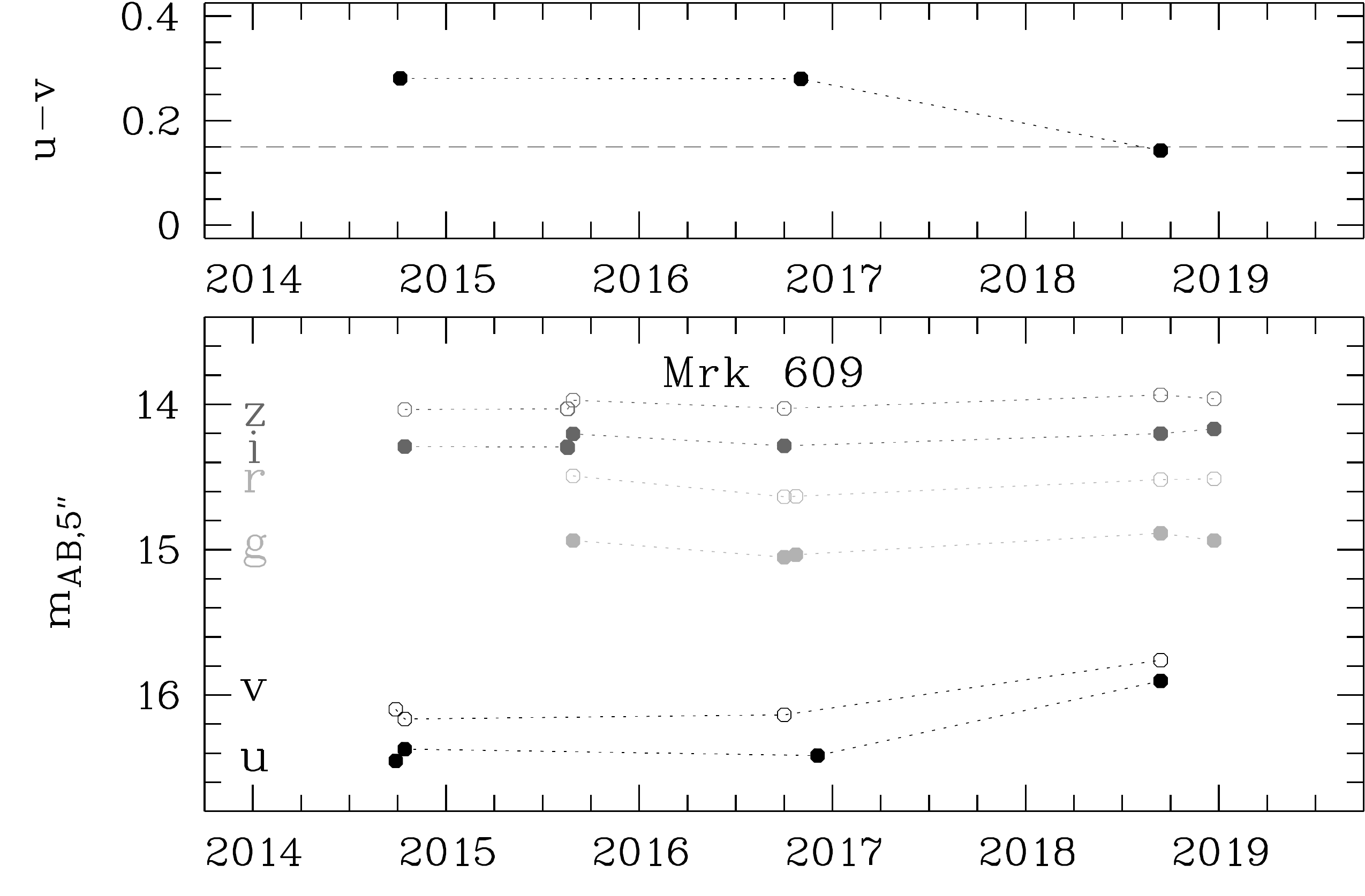} 
\includegraphics[angle=270,clip=true,width=0.975\columnwidth]{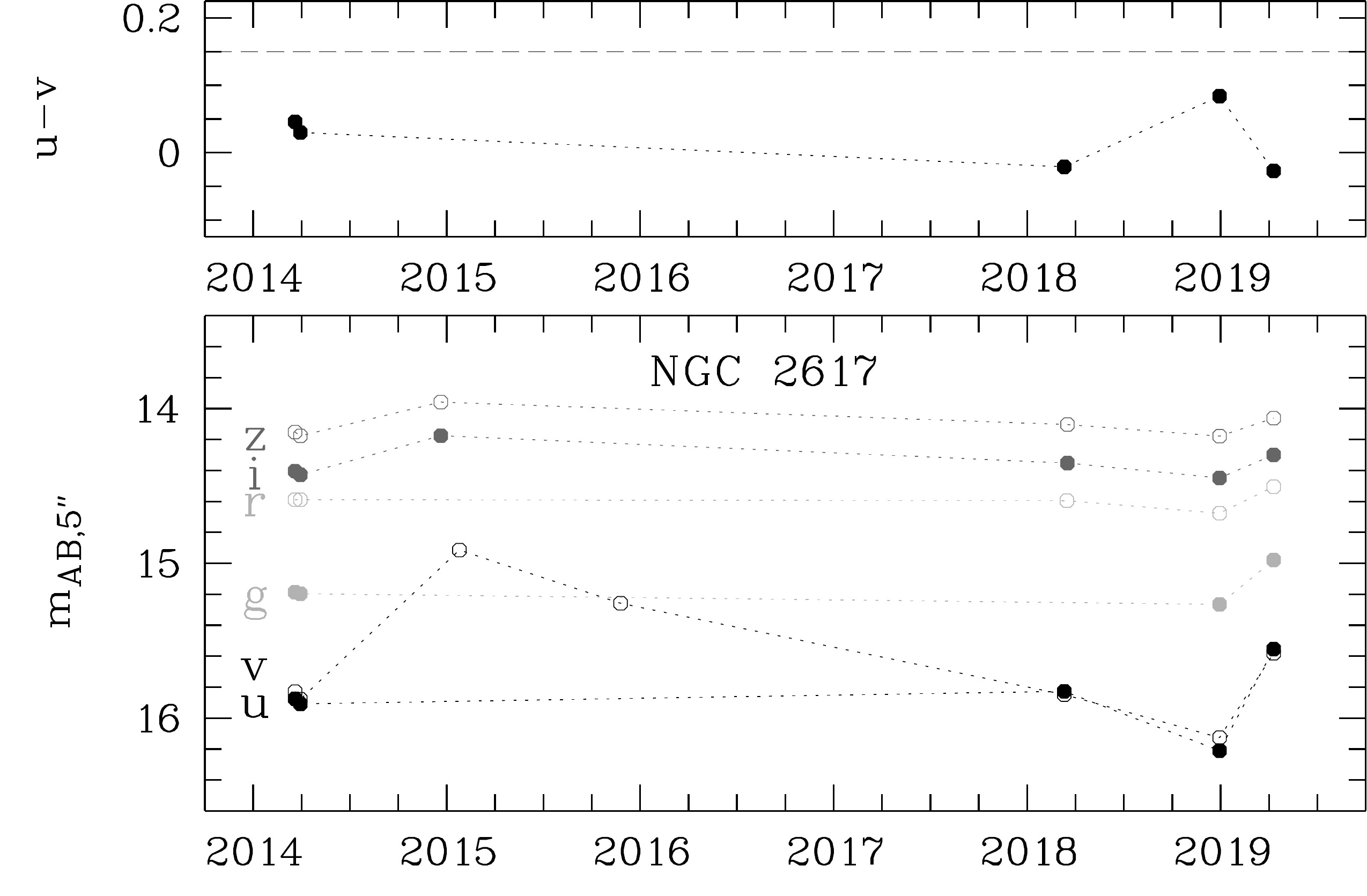} 
\includegraphics[angle=270,clip=true,width=0.975\columnwidth]{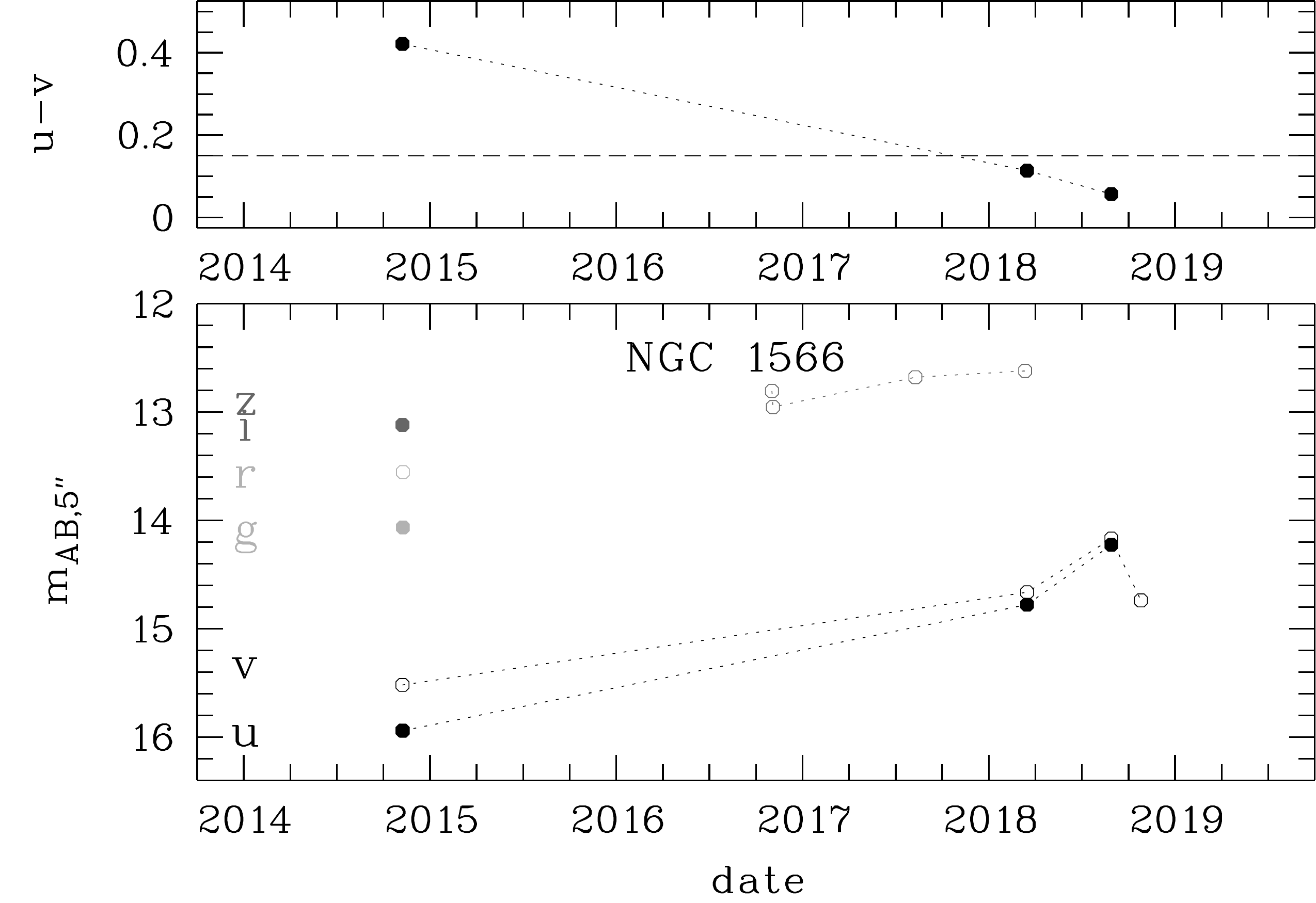} 
\caption{SkyMapper light curves (lower panels) and $u-v$ evolution (upper panels) of CLAGN that show significant variability in the DR3 data set. {\refbf Error bars are smaller than the size of the points.} The dashed line is the selection cut for broad-line AGN in Eq.~\ref{selcut}.
}\label{LCs}
\end{center}
\end{figure}

Fig.~\ref{CLD_CL} shows the path of the known CLAGN in the plane of colour vs. nuclear brightness, and Fig.~\ref{LCs} shows the light curves of the objects with the strongest changes. {\refbf We note in particular, how the CL-AGN with larger variability nicely track the expected orientation of mixing sequences shown in Fig.~\ref{CLD4}.} 

In the following, we compile the scattered notes from the literature on the changes observed in individual objects and discuss how they compare with the observations {\refbf included in DR3 of} the SkyMapper Survey: \\

{\noindent \bf NGC~7582:} \citet{Aretxaga99} reported the appearance of a nuclear type-1 spectrum during late June or early July 1998, which developed through a type 1.8/1.9 in October of the same year. \citet{Aretxaga99} discuss that the galaxy had previously been a classical Seyfert-2 galaxy, with outflows and IR colours suggesting a thick dust torus and an obscured type-1 AGN, which they found hard to reconcile with the type transition. From X-ray data, \citet{Bianchi09} argue that the geometry of this AGN is complex with the dusty torus not being along the line of sight and additional absorbing material causing short-term variability. Finally, \citet{Ricci18} detect at the location of the nucleus a broad H$\alpha$ line in Gemini/GMOS spectra from 2004 and a broad Br~$\gamma$ line in VLT/SINFONI spectra taken in 2007. 

If any type-1 episodes are as short-lived as the case in 1998, SkyMapper is likely to miss them, as the overall number of repeat visits is small; the Main Survey schedule plans visits in the $uv$ filter pair only on two nights for any field. We still see modest variability in SkyMapper DR3, notably a decline of nearly 0.2~mag in $u$ and $v$ bands from 28 June 2018 to 18 Sep 2018 and a $\sim 0.1$~mag rise within the 24 hours to 19 Sep 2018. However, with $u-v\approx 0.4$ throughout the variations, we would expect the AGN to have been a type-1.9 or 2 at all observed epochs. 
\vspace{2mm}

{\noindent \bf Mrk~1018:} This galaxy has been known as a poster-child CLAGN since it first turned from a Sy-1.9 into a Sy-1 in 1984 \citep{Cohen86}. By 2015 it had turned back into a Sy-1.9 \citep{McElroy16}. 

The fairly faint object ($u_{\rm APC05}\approx 18$) was observed by the SkyMapper Main Survey in $u$ and $v$ only on 29 October 2014. The $riz$ bands all {\refbf appear to} show a brightening by 0.14~mag on 10 August 2014 relative to 8 and 11 August 2014, {\refbf however, these observations occurred in untypical seeing and are probably insignificant}. \citet{Krumpe17} report that Mrk 1018 reached a minimum around October 2016 and rebrightened in 2017. DR3 has further visits in $i$ and $z$ band in 2018 and 2019, showing that Mrk 1018 brightened by $\Delta i\approx 0.17$ from 31 July 2015 to 12 August 2018 and stayed at that higher level also on 23 Sep 2019.
\vspace{2mm}

{\noindent \bf NGC~863=Mrk~590:} \citet{Denney14} summarise the spectral evolution of this galaxy over 40 years until 2014. In this period, a peak of activity was seen around 1989 with Mrk~590 in a type-1 state, which was followed by a period of high variability and a decline via type~1.8 in 2003 and 1.9 in 2006 into a type~2 AGN in 2013. 
\citet{Mathur18} argued that the AGN reawakened by November 2014 based on observations of broad Mg{\sc ii} and C{\sc iii} emission lines, although no optical spectrum was available to comment on the Hydrogen lines. 

The SkyMapper DR3 photometry shows modest variability in brightness from August 2014 to July 2015. By Aug/Sep 2018 the galaxy had brightened in all bands, from $\Delta z=-0.13$ to $\Delta u=-0.50$ and all of its colours became bluer with $u-v$ changing from $+0.26$ in Oct 2014 to $+0.19$ in 2018. Compared to other Seyfert galaxies in our sample, this makes us suspect that during 2014/5 a type between 1.8 and 2 was most likely, while by 2018 the object had come close to the Sy-1 selection in Eq.~\ref{selcut}. 
\vspace{2mm}

{\noindent \bf NGC~7603=Mrk~530:} \citet{TO76} first reported a dramatic disappearance of the H$\beta$ line in this type-1 AGN between Nov 1974 and Nov 1975, suggesting a temporary type-1.9/2 phase, and by 1976 the line had started to reappear. \citet{Kollatschny00} discussed its generally strong variability over 20 years of spectral monitoring from 1979 to 1998. By 2006 the object was still seen as a type-1 by \citet{Trippe10}. 

The SkyMapper DR3 photometry shows that Mrk~530 rose by $\sim 0.4$~mag in $u$ band from 2-7 July 2014 to 1 July 2015 (see Fig.~\ref{LCs}); at the same time its $u-v$ colour changed from $0.0$ to $-0.10$. The $riz$ light curve shows that a high state occurred on 27 June only ten days before the lower state in July. We see another dip in September 2017 and another rise in July 2018. The colours suggest that the galaxy has been a type-1 from June 2014 to 2018.
\vspace{2mm}

{\noindent \bf Mrk~609:} according to \citet{Rudy88} this galaxy had been a type-1.8 in 1976, but the broad Hydrogen lines had strengthened and changed ratio by 1984. \citet{Trippe10} quote that this galaxy has been seen as a type-1.5 in the past, perhaps referring to 1984 period, while their observations in 2002 and 2007 as well as SDSS and 6dFGS spectra from 2001 and 2004, respectively, show a type-2 AGN. 

In SkyMapper DR3 we see significant variability (see Fig.~\ref{LCs}), with two low states in October 2014 and October 2016 showing $u-v=0.28$ in each case, and two high states, a brief one during late August 2015 and a possibly enduring one from August 2017 to September and December 2018, with $u-v\approx 0.14$. This makes the galaxy a candidate for changing between type-2 in 2014 to maybe type-1 to 1.5 in 2015, back to type-2 in 2016 and back again to type-1 to 1.5 in 2017 and 2018. However, a spectrum by \citet{Senarath20} from 14 September 2018 shows it as a type-1.9.
\vspace{2mm}

{\noindent \bf NGC~1346:} This galaxy was previously classified as a Sy-1 AGN \citep{VCV03}, but as  \citet{Senarath19} report, an SDSS spectrum from 2001 shows it as a type-1.8/1.9 and three later spectra all show it as a type-2: the 6dFGS spectrum from 2004, a spectrum in 2013 from \citet{Thomas17} and their own spectrum from 12 December 2018. 

SkyMapper DR3 has only one epoch for the $u-v$ colour, which was $+0.34$ on 29 November 2016, clearly consistent with a type-2 AGN. The photometry in the other bands shows little variability, with one exception: from January to August 2015 the galaxy brightened by $(\Delta i,\Delta z) =(-0.17,-0.11)$~mag, and by December 2016 it had fallen back to the level of January, where it was still seen in September 2019. This may suggest an epoch of activity in late 2015 that could have been accompanied by broad lines.
\vspace{2mm}

{\noindent \bf NGC~2617:} \citet{Shappee14} report a change from type-1.8 in 2003 to type-1 by 2013. \citet{Oknyansky17} report further variability and argue that it has been a type-1 since 2012 and until their last observations in 2016. \citet{Oknyansky18} mention that their monitoring has detected a low state from April 2017 until May 2018; a photometric minimum was observed in December 2017 and again in April/May 2018, where a spectrum showed only 'a weak H$\beta$ line'. Afterwards, the galaxy brightened again in 2018 and \citet{Senarath20} took a spectrum on 8 January 2019 showing the object had returned to type-1.

The SkyMapper data (see Fig.~\ref{LCs}) show significant variability from 2014 to 2019. As the source faded by $\Delta u\approx 0.4$ from 11 March 2018 to 29 December 2018, its $u-v$ colour reddened from $-0.03$ to $+0.08$. However, by 10 April 2019 it had brightened by $\Delta u\approx -0.65$ and resumed its bluer level with $u-v-0.03$ again. The $griz$ band light curves have their lowest minimum in December 2018, at which stage we would still expect it to have been a type-1-1.5 based on the $u-v$ colour. The highest state within the coverage of DR3 was reached in January 2015, when the source was $\Delta v=-1.21$~mag higher than during the minimum. There might be a mild tension between the blue $u-v$ colour on 11 March 2018 and the 'weak H$\beta$ line' seen by \citet{Oknyansky18} in a spectrum from 'April/May 2018'.
\vspace{2mm}

{\noindent \bf NGC~1566:} this galaxy is known for several recurrent episodes. \citet{Alloin86} report four outbursts with rise times as short as 20~days for the change from type~1.9 to type~1, followed by exponential decays with $\tau \approx 400$~days. \citet{Peterson88} and later \citet{Oknyansky19} summarise the historical variations in this object that were noticed as early as in the 1950s. Concerning the period of the SkyMapper survey observations, \citet{Oknyansky19} report that a low state was observed from 2014 to early 2017. From September 2017 the source started to rise and when it was observed on 6 July 2018 after a break it was seen at a much brighter level from which it quickly declined over the following two months of observations. Their spectrum taken on 2 August 2018 was reported to suggest a Sy-1.2 type.

The SkyMapper data (see Fig.~\ref{LCs}) show the source in a low in 2014 with a colour of $u-v=0.42$ indicative of a Sy-1.9/2 AGN. By 16 February 2018 it had brightened by $\sim 1$~mag in $u$ band and reached $u-v=0.19$ getting close to our Sy-1 selection cut. By 16 March 2018 it had brightened by a further $\sim 0.2$~mag in $u$ band and reached $u-v=0.11$ making it a strong Sy-1 to 1.5 candidate. The next SkyMapper $uv$ observations were from 28 August 2018, in the declining phase after the peak observed by \citet{Oknyansky19}: the SkyMapper $u$ band was still brighter by an additional $\sim 0.5$~mag relative to March 2018 and the $u-v$ colour had reached $+0.06$, entirely consistent with a Sy-1.2 AGN. Two months later it had declined in $v$ band to the level in Feb/Mar 2018.
\vspace{2mm}

All the above variability from SkyMapper is consistent with nuclear-only changes in brightness, because on an absolute scale the Petrosian flux changes roughly in line with the nuclear aperture flux. The correlated changes in colour and brightness as well as the correlations between $u-v$ index and spectral types emphasise that the SkyMapper $u-v$ index is a useful proxy for Seyfert types. Monitoring AGN with the SkyMapper $u$ and $v$ bands is thus useful for detecting variability in AGN that goes beyond regular variability into the regime of type changes. Deciding whether CLAGN mark simply an extreme tail in a single mode of AGN variability, or constitute a separate phenomenon appearing in a specific subset of AGN, will require larger statistics of AGN variability.

\section{Contaminating transients?}

{ \refbf Nuclear variability originates not always from a continuously accreting black hole. Additional light may also appear temporarily from transients such as tidal disruption events \citep[TDE; see e.g. review by][]{Komossa15} or from supernovae in the nuclear region of the galaxy. }

\subsection{Tidal-disruption events}\label{TDE}

{ \refbf
TDEs produce blue featureless continua as much as AGN do, although their line emission may differentiate them from AGN, involving not only Balmer lines but He II emission and possibly Ne III emission resulting from Bowen flourescence, and by showing evolution as the TDE progresses \citep[see e.g.][]{Arcavi14,Blagorodnova19,Leloudas19}. As such, we expect that we will have no means to photometrically distinguish TDEs and CL-AGN based on only a single epoch of photometry. With multiple epochs, however, the two cases could be distinguished as TDEs usually fade back to nearly host level after less than 200~days in the SkyMapper $u$ band at $z\la 0.1$ \citep[see example light curves in][]{Holoien18, Leloudas19, Holoien20}. 

\citet{Auchettl18} estimate the rate of TDEs to be 70 to 470 per million years per galaxy. If these were detectable for (optimistically) half a year in the SkyMapper $u-v$ colour, we might have between 1 and 6 TDEs in the sample of this paper. They could thus be a modest but noticeable portion among the 46 plus 16 candidates selected by Eq.~3, which have archival spectra showing type-1.8 to 2 AGN and normal galaxies, respectively. Correlating known TDEs with the SkyMapper data set will be reserved for a separate paper.}

\begin{figure}
\begin{center}
\centering
\includegraphics[width=\columnwidth]{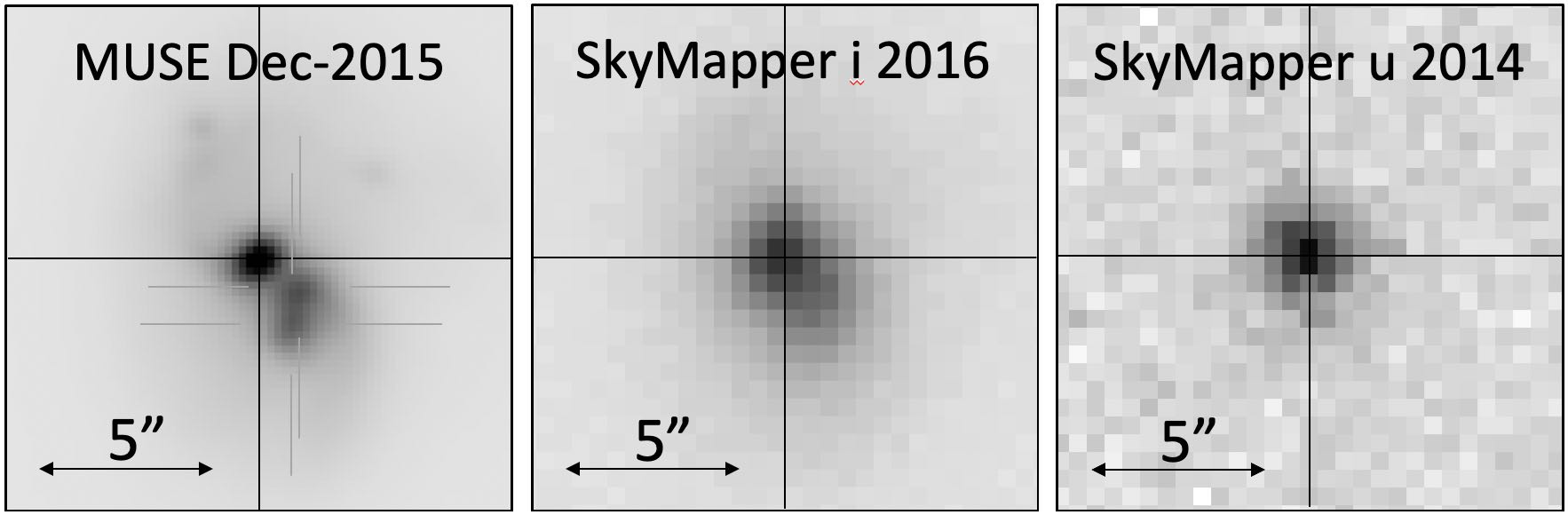}
\includegraphics[angle=270,width=\columnwidth]{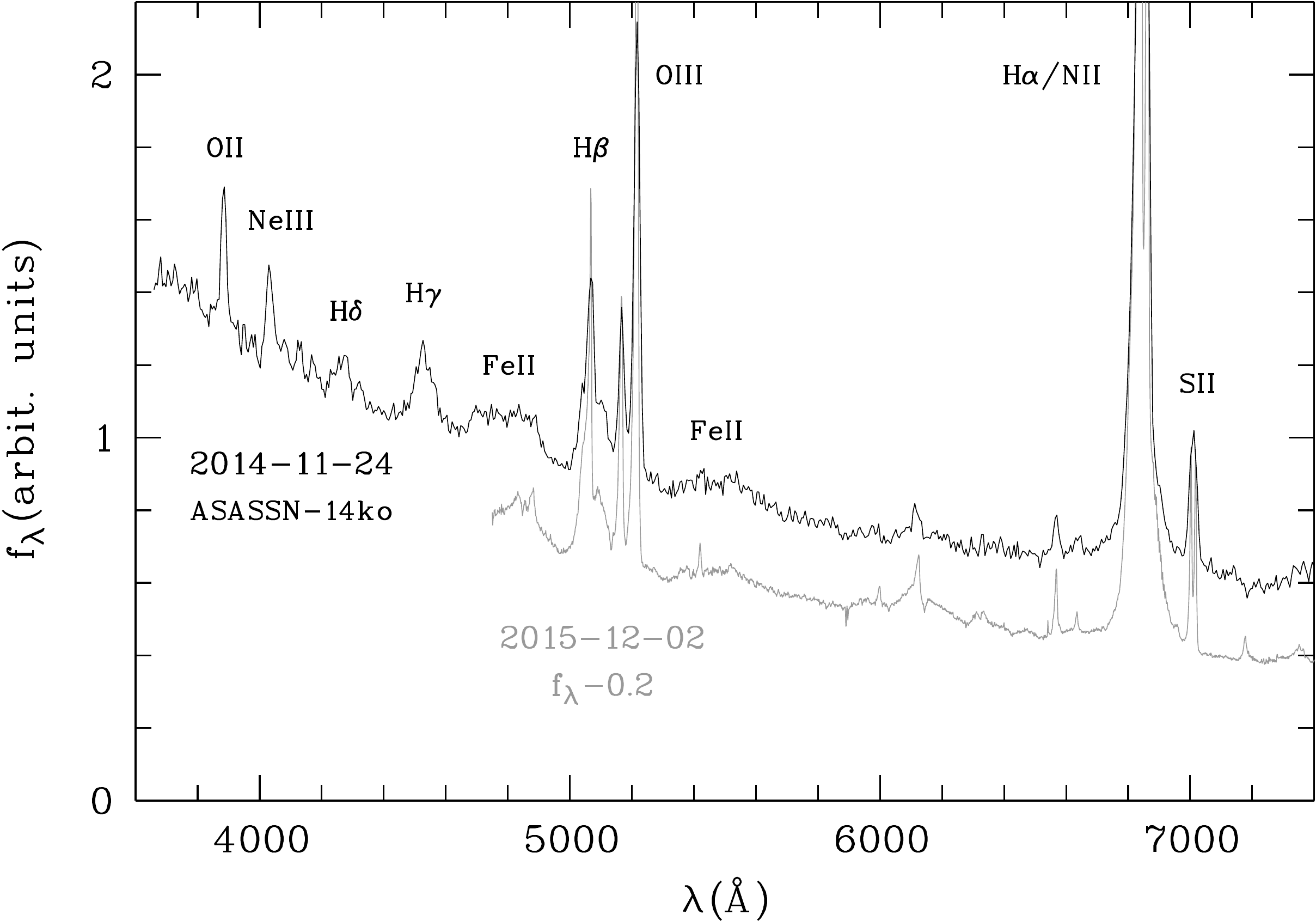} 
\caption{ {\refbf 
Images of ESO253-G003 (top): VLT/MUSE shows three nuclei; SkyMapper $u$ band on 9 Nov 2014 relative to $i$ band in 2016 shows the transient ASASSN-14ko centred on the Northeast (top left) nucleus. Spectra of ASASSN-14ko at $z=0.0425$ (bottom) by \citet{Dimitriadis14} and of the bright Northeast nucleus by \citet{2020AJ....159..167L}. }
}\label{specs_14ko}
\end{center}
\end{figure}

\subsection{Supernovae}\label{14ko}

{ \refbf
For supernovae, we estimate colours from template spectra of different supernova types \citep[e.g.][]{Filippenko97,Foley16} at the redshifts of $z\la 0.1$ considered here and find that their SkyMapper $u-v$ colours are redder than AGN continua due to metal absorption bands. This appears to be the case for supernovae of all common types including Ia, Ic and II, and irrespective of the fact that the SkyMapper $v-g$ and $g-i$ colours can successfully differentiate thermonuclear SNe Ia from core-collapse SNe Ic and II at these redshifts \citep{Scalzo17}.

We find one particular entry in the Open Supernova Catalog \citep[OSC,][]{OSC}, which is one of our turn-on CL-AGN candidates and also has $u-v$ data around its maximum: ASASSN-14ko is listed as a type IIn supernova discovered on 14 Nov 2014 by \citet{Holoien14} about $0\farcs 85$ from the centre of the galaxy ESO 253-G003 at $z=0.0425$ that has been typed as a Seyfert-2 galaxy from an archival spectrum. In the following, we attempt to establish whether this object is indeed a contaminating supernova or a type-1 AGN as suggested by our $u-v$ colour. 

SkyMapper $uv$ photometry covers the object on 9+10 Nov 2014, 11 Mar 2018, 15 Aug 2019 and 8 Oct 2019. The object has a bright state with $(u,v)=(15.86\pm 0.01,15.82\pm 0.01)$ that corresponds to the possible SN peak in Nov 2014 and again with $(u,v)=(15.87\pm 0.01,15.80\pm 0.01)$ in Oct 2019. In between it is seen in a fainter state with $(u,v)=(16.60\pm 0.04,16.46\pm 0.07)$ in Mar 2018 and $(u,v)=(16.84\pm 0.05,16.75\pm 0.04)$ in Aug 2019. Based on the $u-v$ colour, ranging from $\sim 0.05$ in the bright state to $\sim 0.12$ in the faint state, we considered this object a candidate Seyfert-1 galaxy, and based on the earlier Seyfert-2 classification it is one of our 46 new candidate turn-on Changing-Look AGN. 

Usually, AGN are not expected to occur $0\farcs 85$ ($\sim 700$~pc in projection) away from the nucleus of a galaxy. However, the SkyMapper $i$ band images reveal that this object is a post-merger galaxy surrounded by a tidal arm; we see two nuclei with $\sim 1\farcs 5$ separation, and the $u$ band image of Nov 2014 shows ASASSN-14ko as a bright point source precisely at the location of the North-Eastern nucleus (see Fig.~\ref{specs_14ko}). ASASSN-14ko is thus a nuclear transient and may just be a flare of an AGN. Recently published VLT/MUSE observations by \citet{2020AJ....159..167L} reveal even three nuclei, as the Southwestern nucleus is resolved into two cores (see Fig.~\ref{specs_14ko}). 

\citet{Holoien14} mention in their discovery telegram that {\it strong AGN variability, while unusual for Type II AGN, cannot be ruled out}. \citet{Dimitriadis14} label it {\it consistent with a SN IIn} from a spectrum taken on 24 Nov 2014 at the ESO/NTT. We reproduce the spectrum\footnote{from \url{https://wiserep.weizmann.ac.il/object/5901}} in Fig.~\ref{specs_14ko} but do not recognise any unique SN features; to our eyes, it is indistinguishable from the SDSS-based Seyfert-1 template by \citet{Pol17}. 
The MUSE data by \citet{2020AJ....159..167L} were taken in December 2015 and show a complex source structure. 
Importantly, broad Balmer lines are seen at the location of the North-Eastern nucleus, which appear 
similar to the earlier NTT data, more than a year after the initial ASASSN transient. 
Since the blue $u-v$ colour stretches at least five years, we claim that the transient ASASSN-14ko is not a supernova but the flaring tip of the iceberg of a highly variable, newly discovered Changing-Look AGN. Finally, \citet{Payne20} argue in a new paper, that this object is even a periodically flaring AGN. }

\section{Conclusions}

In this paper, we investigate the nuclear colours of low-redshift galaxies and AGN, at $z\la 0.1$, using photometry in the SkyMapper Southern Survey Data Release 3. We place particular emphasis on the bespoke ultraviolet $u$ and violet $v$ filters in the SkyMapper filter set, which are centred on $\sim 350$~nm and $\sim 385$~nm, respectively.

We find that the $u-v$ colour index is effectively constant for all galaxies undergoing secular evolution, irrespective of the age of a stellar population. Instead, $u-v$ is an indicator of recent change in star formation rate and responds thus to starbursts and quenching events. We also find that type-1 AGN are well differentiated from inactive galaxies. Type-1 AGN are the bluest galaxies in $u-v$, covering mostly the range of $[-0.2,+0.15]$, while starburst galaxies range across $[+0.1,+0.3]$; galaxies with only secular changes in star-formation rate are centred on $u-v=0.4$, while galaxies undergoing quenching are expected to be found at $u-v>0.5$. Using SEDs from population synthesis we establish mixing sequences from pure starlight to luminous unobscured AGN.

From SkyMapper photometry of over 25,000 nearby galaxies in the 2MRS and 6dFGS catalogues as well as 688 spectral types that we eyeballed in the 6dFGS atlas, we develop a selection rule for type-1 AGN that uses a combination of $u-v$ colour and nuclear brightness. Objects that do not obey the rule are expected to be either type-2 AGN or inactive galaxies. This selection is expected to be sensitive to Seyfert-1 nuclei with luminosities {\refbf as low as} $M_B\approx -15.5$ at $z=0.01$ {\refbf and brighter than} $M_B\approx -19.5$ at $z=0.08$.

We notice several galaxies whose spectral type seen by 6dFGS appears to disagree with the type predicted from our selection rule and SkyMapper photometry. Some of these are known Changing-Look AGN (CLAGN) that have changed their spectral AGN type over the course of recent years, and plausibly between the epoch of 6dFGS spectroscopy and the epoch of SkyMapper photometry. Two of these CLAGN even move across the selection line during the five years of repeat observations with SkyMapper and have likely changed their spectral type alongside.

A larger number of such type disagreements have not yet been reported as CLAGN, but we speculate that several of these may be CLAGN. {\refbf One of these objects in particular, known previously as the transient ASASSN-14ko is revealed as a new Changing-Look AGN. The nature of further CL-AGN candidates will be studied in a forthcoming paper by Hon et al. (in preparation). Some of the shorter-lived $uv$-blue phases may reveal tidal-disruption events, which we statistically expect to see in this sample as well. 

Two issues limit the utility of the $u-v$ method: (1) the $u-v$ colour does not detect the emission lines that underlie the Seyfert classification scheme; instead, it detects the blue featureless continuum in AGN, which is mostly but not always correlated with the visibility of broad emission lines. (2) The} mixing sequence from starlight to type-1 AGN is continuous and depends on the luminosity of the nucleus relative to the nuclear stellar population. Since the $u-v$ colour intertwines AGN type and AGN luminosity contrast, there is genuine overlap between the AGN types in terms of nuclear $u-v$ colours. Sorting out what is genuine overlap in type locus and what are evolutionary changes in AGN type requires spectra during the epoch of the photometric observations. While the photometry of SkyMapper DR3 is already a few years in the past, current spectra could still find evidence of persistent changes.

\section*{acknowledgements}
We thank an anonymous referee for suggestions improving the manuscript, Katie Auchettl for discussion, and Mark Krumholz for advice on using his population synthesis code SLUG. 
This work was partly supported by the Australian Research Council Centre of Excellence for All-sky Astrophysics (CAASTRO), through project number CE110001020. JG acknowledges support from a CAASTRO pre-PhD project in contributing to this work. WJH would like to thank the University of Melbourne for an International Graduate Research Scholarship. CAO acknowledges support from the Australian Research Council through Discovery Project DP190100252.
The national facility capability for SkyMapper has been funded through ARC LIEF grant LE130100104 from the Australian Research Council, awarded to the University of Sydney, the Australian National University, Swinburne University of Technology, the University of Queensland, the University of Western Australia, the University of Melbourne, Curtin University of Technology, Monash University and the Australian Astronomical Observatory. SkyMapper is owned and operated by The Australian National University's Research School of Astronomy and Astrophysics. The survey data were processed and provided by the SkyMapper Team at ANU. The SkyMapper node of the All-Sky Virtual Observatory (ASVO) is hosted at the National Computational Infrastructure (NCI). Development and support of the SkyMapper node of the ASVO has been funded in part by Astronomy Australia Limited (AAL) and the Australian Government through the Commonwealth's Education Investment Fund (EIF) and National Collaborative Research Infrastructure Strategy (NCRIS), particularly the National eResearch Collaboration Tools and Resources (NeCTAR) and the Australian National Data Service Projects (ANDS). This research has made use of the SIMBAD database, operated at CDS, Strasbourg, France.

\section*{DATA AVAILABILITY}
The SkyMapper data underlying this article are available at the SkyMapper node of the All-Sky Virtual Observatory (ASVO), hosted at the National Computational Infrastructure (NCI) at \url{skymapper.anu.edu.au}. The data from Data Release 3 are currently accessible only to Australia-based researchers and their collaborators.

\bibliographystyle{pasa-mnras}

\appendix

\section{Additional tables}

\begin{table*}
\centering          
\caption{Broad-line AGN (type-1 to 1.8) Newly identified in this work by eyeballing 6dFGS spectra.}
\label{newSy1tab}      
\begin{tabular}{Hlrrcclr}
\hline       
id & 6dFGS name & RA & Dec & Redshift & 6dFGS type & Simbad type & SkyMapper ID  \\ 
\hline
        1 & g0000439$-$260522 &  0.1830 &$-$26.0892 &  0.059   &  1.0 & Star             &  11447449 \\
        4 & g0016379$-$054425 &  4.1580  &$-$5.7402 &  0.074   &  1.5 & AGN              &  13137963 \\
        9 & g0040392$-$371317 & 10.1632 &$-$37.2213 &  0.036   &  1.0 & Galaxy            &  9103187 \\
       14 & g0115274$-$312352 & 18.8642 &$-$31.3978 &  0.053   &  1.5 & Galaxy           &  10044198 \\
       15 & g0119066$-$325713 & 19.7777 &$-$32.9536 &  0.067   &  1.5 & Galaxy            &  9995395 \\
       22 & g0225363$-$273048 & 36.4014 &$-$27.5133 &  0.064   &  1.0 & AGN\_Candidate    &  11026402 \\
       25 & g0245455$-$030450 & 41.4395  &$-$3.0804 &  0.075   &  1.5 & Galaxy         &  15638721 \\
       26 & g0257503$-$133720 & 44.4594 &$-$13.6221 &  0.079   &  1.0 & AGN\_Candidate    &  14839355 \\
       38 & g0358201$-$053207 & 59.5837  &$-$5.5352 &  0.062   &  1.5 & Galaxy         &  22201714 \\
       41 & g0428400$-$184122 & 67.1664 &$-$18.6896 &  0.070   &  1.8 & AGN\_Candidate    &  22708550 \\
       43 & g0429383$-$210944 & 67.4096 &$-$21.1622 &  0.070   &  1.0 & AGN              &  22460749 \\
       46 & g0435577$-$355858 & 68.9905 &$-$35.9827 &  0.060   &  1.5 & Galaxy         &  18460747 \\
       48 & g0440404$-$411044 & 70.1681 &$-$41.1788 &  0.033   &  1.0 & AGN\_Candidate    &  18336952 \\
       54 & g0527208$-$195411 & 81.8367 &$-$19.9031 &  0.069   &  1.0 & Galaxy         &  23662308 \\
       56 & g0529592$-$340159 & 82.4967 &$-$34.0329 &  0.079   &  1.8 & AGN              &  19795813 \\
       58 & g0551180$-$200548 & 87.8249 &$-$20.0965 &  0.054   &  1.8 & AGN\_Candidate    &  23948916 \\
       61 & g0618453$-$351815 & 94.6886 &$-$35.3041 &  0.045   &  1.8 & EmG              &  31786125 \\
       65 & g0641577$-$431742 &100.4904 &$-$43.2951 &  0.061   &  1.0 & Galaxy           &  28445001 \\
       66 & g0650175$-$380514 &102.5728 &$-$38.0871 &  0.030   &  1.0 & AGN\_Candidate    &  29265277 \\
       69 & g0656543$-$631536 &104.2261 &$-$63.2598 &  0.035   &  1.0 & Galaxy          &  464729086 \\
       73 & g0802539$-$582137 &120.7248 &$-$58.3604 &  0.060   &  1.5 & AGN\_Candidate   &  467574241 \\
       75 & g0827341$-$021247 &126.8923  &$-$2.2130 &  0.039   &  1.5 & Galaxy          &   75828917 \\
       80 & g0941115$-$112822 &145.2981 &$-$11.4729 &  0.047   &  1.5 & Galaxy           &  89665239 \\
       81 & g0945126$-$123632 &146.3024 &$-$12.6088 &  0.048   &  1.0 & EmG              &  88896854 \\
       82 & g0954025$-$341343 &148.5106 &$-$34.2285 &  0.032   &  1.0 & EmG              &  80229823 \\
       85 & g1014114$-$285631 &153.5477 &$-$28.9420 &  0.065   &  1.0 & Galaxy           &  85819989 \\
       90 & g1047457$-$375932 &161.9405 &$-$37.9922 &  0.075   &  1.0 & Galaxy           &  83561935 \\
       91 & g1048338$-$390238 &162.1410 &$-$39.0439 &  0.045   &  1.0 & AGN              &  83509041 \\
       92 & g1053410$-$291824 &163.4208 &$-$29.3067 &  0.058   &  1.0 & Galaxy           &  86272917 \\
       96 & g1109415$-$033927 &167.4230  &$-$3.6573 &  0.039   &  1.5 & Galaxy           &  93579464 \\
      109 & g1248211$-$272550 &192.0880 &$-$27.4306 &  0.066   &  1.0 & AGN\_Candidate    &  98131214 \\
      110 & g1250189$-$233358 &192.5788 &$-$23.5660 &  0.048   &  1.0 & AGN             &  104071624 \\
      114 & g1328311$-$490906 &202.1296 &$-$49.1516 &  0.048   &  1.5 & AGN\_Candidate   &  320346080 \\
      116 & g1343431$-$395032 &205.9297 &$-$39.8422 &  0.056   &  1.5 & AGN\_Candidate    &  95998184 \\
      118 & g1353296$-$434143 &208.3735 &$-$43.6954 &  0.038   &  1.0 & Galaxy          &  318670924 \\
      123 & g1406507$-$244250 &211.7114 &$-$24.7138 &  0.046   &  1.8 & AGN             &  102488645 \\
      124 & g1407220$-$312058 &211.8417 &$-$31.3493 &  0.075   &  1.0 & AGN\_Candidate   &  101746230 \\
      129 & g1423502$-$092318 &215.9594  &$-$9.3881 &  0.068   &  1.0 & AGN\_Candidate   &  109988713 \\
      132 & g1437457$-$271803 &219.4403 &$-$27.3009 &  0.065   &  1.0 & AGN\_Candidate   &  103030136 \\
      142 & g1530395$-$141712 &232.6646 &$-$14.2865 &  0.052   &  1.0 & AGN\_Candidate   &  166847177 \\
      143 & g1552438$-$165341 &238.1824 &$-$16.8947 &  0.024   &  1.5 & Galaxy          &  168244857 \\
      144 & g1611435$-$343529 &242.9314 &$-$34.5912 &  0.061   &  1.5 & Galaxy          &  117747295 \\
      146 & g1623189$-$301433 &245.8289 &$-$30.2426 &  0.060   &  1.5 & Galaxy          &  141486947 \\
      148 & g1639001$-$012857 &249.7502  &$-$1.4826 &  0.091   &  1.0 & AGN\_Candidate   &  194846346 \\
      151 & g1729163$-$685442 &279.8993 &$-$35.6899 &  0.046   &  1.0 & AGN\_Candidate   &  216993546 \\
      154 & g1839358$-$354124 &279.8993 &$-$35.6899 &  0.046   &  1.0 & AGN\_Candidate   &  216993546 \\
      165 & g2022088$-$471452 &305.5368 &$-$47.2475 &  0.031   &  1.8 & Galaxy          &  487162588 \\
      166 & g2025574$-$482226 &317.6383 &$-$52.4621 &  0.071   &  1.0 & Galaxy          &  487661480 \\
      171 & g2110332$-$522744 &317.6383 &$-$52.4621 &  0.071   &  1.0 & Galaxy          &  487661480 \\
      175 & g2147133$-$120126 &326.8053 &$-$12.0238 &  0.048   &  1.5 & Galaxy            &  3971217 \\
      180 & g2218212$-$300448 &334.5880 &$-$30.0798 &  0.070   &  1.0 & AGN               &  2061541 \\
\hline                  
\end{tabular}
\end{table*}

\end{document}